\documentclass[structabstract]{aa}  

\usepackage{graphicx}
\usepackage{txfonts}
\usepackage{natbib}
\usepackage{color}

\usepackage{caption}
\usepackage{subfig}
\usepackage{hyperref}

\def\asec{\ifmmode ^{\prime\prime}\else$^{\prime\prime}$\fi}

\def\Ms{\mbox{\,M$_{\odot}$}}

\def\rsun{\hbox{R$_{\odot}$}}

\def\degs{\ifmmode ^{\circ}\else$^{\circ}$\fi}
\def\amin{\ifmmode ^{\prime}\else$^{\prime}$\fi}
\def\asec{\ifmmode ^{\prime\prime}\else$^{\prime\prime}$\fi}
\def\farcs{\hbox{$.\!\!^{\prime\prime}$}}  % Fractions of arcseconds
\def\degs{\ifmmode ^{\circ}\else$^{\circ}$\fi}
\def\amin{\ifmmode ^{\prime}\else$^{\prime}$\fi}
\def\cm{\mbox{\,cm}}

\def\cm3{\mbox{\,cm$^{-3}$}}
\def\kms{\mbox{\,km~s$^{-1}$}}

\def\kms{\mbox{\,km s$^{-1}$}}

\def\lsim{\!\!\!\phantom{\le}\smash{\buildrel{}\over
 {\lower2.5dd\hbox{$\buildrel{\lower2dd\hbox{$\displaystyle<$}}\over
                                 \sim$}}}\,\,}
\def\gsim{\!\!\!\phantom{\ge}\smash{\buildrel{}\over
{\lower2.5dd\hbox{$\buildrel{\lower2dd\hbox{$\displaystyle>$}}\over
                               \sim$}}}\,\,}

\begin{document}

 \title{Metallicity determination of M dwarfs \thanks{Based on data obtained at ESO-VLT, Paranal Observatory, Chile, Program ID 090.D-0796(A).}  \\ {\Large Expanded parameter range in metallicity and effective temperature} } 
   
\author{Sara Lindgren\inst{1} 
\and Ulrike Heiter\inst{1}
}
          
\institute{$^1$Observational Astrophysics, Department of Physics and Astronomy, Uppsala University, Box 516, 751 20 Uppsala, Sweden\\
{\it \email{sara.lindgren@physics.uu.se}}\\          
           }
\date{Received: 1 March 2017 / Accepted: 23 May 2017}

\abstract
%context
{Reliable metallicity values for late K and M dwarfs are important for studies of the chemical evolution of the Galaxy and advancement of planet formation theory in low-mass environments. Historically the determination of stellar parameters of low-mass stars has been challenging due to the low surface temperature, causing several molecules to form in the photospheric layers. In our work we use the fact that infrared high-resolution spectrographs have opened up a new window for investigating M~dwarfs. This enables us to use similar methods as for warmer solar-like stars. }
%aims
{Metallicity determination using high-resolution spectra is more accurate than the use of low-resolution spectra, but rather time-consuming. In this paper we expand our sample analyzed with this precise method both in metallicity and effective temperature in order to build up a calibration sample for a future revised empirical calibration. }
%methods
{Because of the relatively few molecular lines in the $J$-band, continuum rectification is possible for high-resolution spectra, allowing the stellar parameters to be determined with greater accuracy than using optical spectra. We obtained high-resolution spectra with the CRIRES spectrograph at the VLT. The metallicity was determined using synthetic spectral fitting of several atomic species. For M dwarfs cooler than 3575 K the line strengths of FeH lines were used to determine the effective temperatures, while for warmer stars a photometric calibration was used. }
%results
{We have analyzed sixteen targets, with a range of effective temperature from 3350-4550 K. The resulting metallicities lie between $-0.5$ $<$ [M/H] $<$ +0.4, with the majority of the targets having sub-solar values. A few targets have previously been analyzed using low-resolution spectra, and we find a rather good agreement with our values. A comparison with available photometric calibrations shows varying agreement, and the spread within all empirical calibrations is large. }
%conclusions
{Including the targets from our previous paper, we have analyzed 28 M dwarfs using high-resolution infrared spectra. The targets spread approximately one~dex in metallicity and 1400~K in effective temperature. For individual M dwarfs we achieve uncertainties of 0.05~dex and 100~K on average. }


\authorrunning{Lindgren \& Heiter} 
\titlerunning{Metallicity of M dwarfs}

\maketitle

\section{Introduction}
Low-mass dwarfs (0.08 \Ms~$<$ M $<$ 0.6 \Ms) are, by number the dominant stellar population in the local Galaxy and constitute as much as 70\% of the stars \citep{Covey2008}. Their multitude together with their long lifetime on the main-sequence make them important for studies of the structure and kinematics of stellar populations. These studies require dynamical/chemical evolutionary models that in turn need accurate metallicity values \citep{Bochanski2010, Woolf2012}. 

Despite their intrinsic faintness M dwarfs are becoming attractive targets for exoplanet searches as their smaller mass and radius makes detection of smaller planets easier (e.g. Gaidos et al. 2007). Furthermore, estimates show that smaller planets are more common around these smaller stars \citep{Bonfils2013, Dressing2013, Mulders2015}. The abundances in the photosphere of main-sequence stars are expected to be a good representation of the material from which the star and, if any, planets were formed. Accurate determination of the atmospheric abundances is therefore critical to advance the current understanding of planet formation and explore the planet-host metallicity correlation towards cooler hosts. That FGK dwarfs with giant planets tend to have enhanced metallicity compared to the Sun is today well established (e.g. \citealt{Santos2004, Fischer2005}). Recent studies indicate that a similar metal enhancement is also present among M dwarfs \citep{Johnson2009, Rojas-Ayala2010, Neves2013, Gaidos2014}. The occurrence rate of giant planets around these low-mass stars is likely also scaling with the stellar mass, which may further affect the correlation with metallicity \citep{Johnson2010, Gaidos2013}. Lastly, the determined stellar parameters of the host star have a direct influence on the derived planet properties. For M dwarfs, due to their intrinsic faintness, the radius and mass are typically determined by combining empirical mass-luminosity relationships (e.g. \citealt{Delfosse2000, Mann2015}) with model mass-radius relationships, that in turn depend on the effective temperature and metallicity (e.g. \citealt{Baraffe1998, Feiden2012}).

\begin{table*}
\caption{Basic information for our sample.} 
\label{sampleinfo}
\centering
\begin{tabular}{lcccccc}
\hline\hline
Target 		& RA and Dec (J2000) 			& Spectral type 	& Planet 		& S/N ratio 	& Ref. \\  \hline
GJ~179 		&  04 52 05.731 +06 28 35.64		& M3.5 			& yes		& 150		& 1, 10 \\
GJ~203 		&  05 28 00.153 +09 38 38.14		& M3.5 			& 			&  65			& 1 \\
GJ~228 		& 06 10 54.804 +10 19 04.99 		& M2.5			& 			& 125		& 1 \\
GJ~334 		& 09 06 45.348 $-$08 48 24.61  	& K7 			& 			& 130		& 2 \\
GJ~433 		& 11 35 26.947 $-$32 32 23.90	& M1.5 			& yes		& 125 		& 2, 11 \\
GJ~514 		& 13 29 59.786 +10 22 37.79		& M1 			&			& 90			& 1 \\
GJ~825 		& 21 17 15.269 $-$38 52 02.50	& M1 			& 			& 180 		& 4 \\
GJ~832 		& 21 33 33.975 $-$49 00 32.42	& M1.5 			& yes		& 205 		& 2, 12 \\
GJ~872B 		& 22 46 42.32 +12 10 21.5		& M3 			&			& 100 		& 5 \\
GJ~880 		& 22 56 34.805 +16 33 12.35		& M1.5 			& 			& 175 		& 1 \\
GJ~908 		& 23 49 12.528 +02 24 04.41		& M1.5			& 			& 125 		& 1 \\
GJ~3634 		& 10 58 35.133 $-$31 08 38.29 	& M2.5 			& yes 		& 55 			& 3, 13 \\
GJ~9356 		& 11 17 13.666 $-$01 58 54.67	& K4/K5 			& 			& 125 		& 6 \\
GJ~9415 		& 12 40 46.289 $-$43 33 58.95	& M3 			& 			& 70			& 7 \\
HIP~12961 	& 02 46 42.886 $-$23 05 11.80	& M1 			&			& 95			& 8, 14 \\
HIP~31878 	& 06 39 50.022 $-$61 28 41.53	& K7 			& yes 		& 105		& 9 \\
\hline
\end{tabular}
\noindent
\begin{flushleft}
{\bf References.} (1)~\citet{Lepine2013},  (2)~\citet{Reid1995}, (3)~\citet{Hawley1996}, (4)~\citet{Joy1974}, (5)~\citet{Newton2014}, (6)~\citet{Stephenson1986}, (7)~\citet{Henry2002}, (8)~\citet{Rojas-Ayala2012}, (9)~\citet{Torres2006}, (10) \citet{Howard2010}, (11)~\citet{Delfosse2013}, (12)~\citet{Bailey2009}, (13)~\citet{Bonfils2011}, (14)~\citet{Forveille2011}. 
\end{flushleft}
\end{table*}

Fundamental parameters, such as metallicity, effective temperature and surface gravity for FGK dwarfs are today determined with rather good precision and accuracy using calculations of synthetic spectra or equivalent widths with software like SME \citep{Valenti1996, Piskunov2017} or MOOG \citep{Sneden1973}. Unlike their solar-type counterparts, the stellar parameters of M dwarfs are more challenging to determine. Their low surface temperatures result in plenty of diatomic and triatomic molecules in the photospheric layers. Until recently, high-resolution spectrographs were limited to the visible wavelength region where these molecules give rise to millions of lines. This makes continuum identification nearly impossible, and leaves almost no unblended atomic lines \citep{Gustafsson1989}. 

Most previous studies characterizing M dwarfs have therefore been based on different empirical calibrations. The use of photometric colors to estimate the metallicity was pioneered by \citet{Bonfils2005}, where the star's position in a color-magnitude diagram can be related to the metallicity. This approach has later been revised by \citet{Johnson2009, Schlaufman2010, Neves2012, Johnson2012}. In our previous work \citep[hereafter L16]{Lindgren2016} we showed that for individual stars, different calibrations can give metallicity values that differ by as much as 0.6~dex. In addition to the photometric calibrations, several spectroscopic calibrations have been developed using moderate resolution spectra in the visible \citep{Woolf2006, Woolf2009}, and in the infrared \citep{Rojas-Ayala2010, Rojas-Ayala2012, Terrien2012, Newton2014, Mann2013, Mann2014}, where \citet{Mann2013} also found metallicity sensitive features in the visible. Most of these metallicity sensitive features are either alkaline metals or alkaline earth metals, and therefore not a direct probe of the iron abundance. In addition, they need to account for the pseudo-continuum, that in the infrared mainly stems from H$_2$O whilst it is formed by TiO in the visible. \citet{Veyette2016} showed in a recent paper that the C/O ratio has a strong influence on the strength of these molecular bands, and that the metallicity determined using spectroscopic calibrations is influenced by the C and O abundances.

In our work we use the fact that high-resolution spectrographs in the infrared have opened up a new window for investigating M dwarfs. In the $J$-band the number of molecular transitions is greatly reduced, limiting the amount of blends with atomic lines and allowing a continuum rectification. This enabled us to use similar methods as is standard for warmer solar-like stars in the visible to determine the overall metallicity through synthetic spectral fitting. We have verified the reliability of our method by analyzing both components in several M+FGK binaries (\citealt{Onehag2012}; L16), showing that we achieve very similar metallicities for both components, with estimated uncertainties of 0.1~dex on average for the M dwarfs. Obtaining and analyzing high-resolution spectra of M dwarfs is time-consuming. Thus for an analysis of large samples of hundreds or thousands of M dwarfs, which are needed for dynamical and chemical Galactic studies, our approach may not be optimal. We therefore set out to explore the possibility of a new empirical calibration based on metallicities of individual M dwarfs. Previous calibrations have all used M dwarfs in binaries where the metallicity was taken from the solar-like component. The previous analysis of M dwarfs in L16 focused on a few binaries and known planet hosts, resulting in a rather metal-rich sample and with a small range in effective temperature. In order to have a good basis for a future calibration, we need to expand the sample both in metallicity and effective temperature.

\begin{figure*}
\centering
\includegraphics[width=0.9\textwidth]{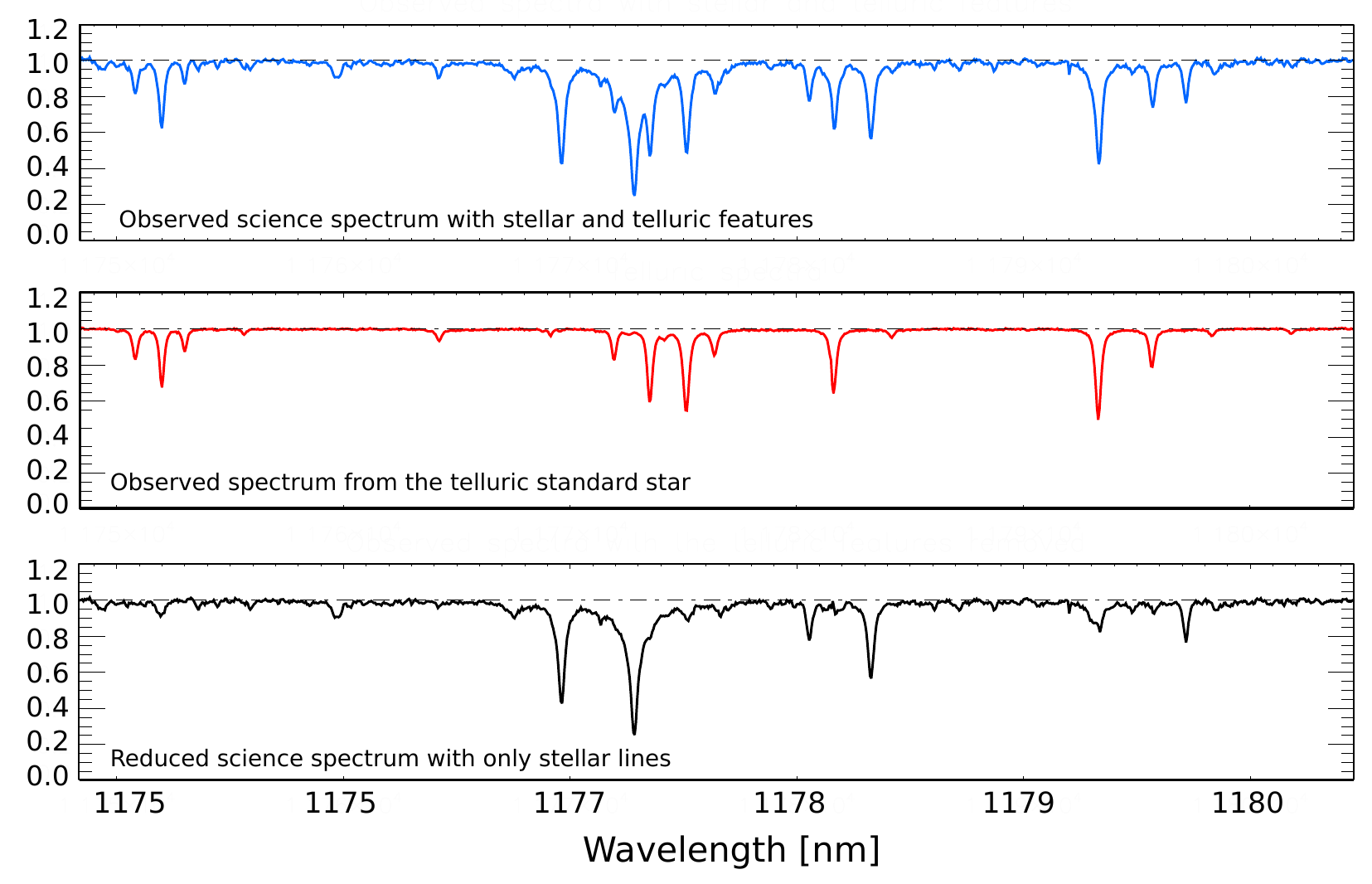} 
\caption{An example of a continuum rectified and wavelength corrected spectrum in a short wavelength interval for target GJ~179. The top panel shows the science spectrum, the middle panel the spectrum from the telluric standard star and the bottom panel the science spectrum where the telluric lines have been removed. We added a horizontal dashed line at one for visualization of continuum regions.
\label{example} }
\end{figure*}

\section{Sample selection and observations}
The targets were selected based on metallicity estimates from the photometric empirical calibration by \citet{Schlaufman2010}, with the goal to include several targets with sub-solar metallicity. The selection was furthermore based on targets with parallaxes $>$25~mas with uncertainties less than 5\%. Our observed sample includes 22 targets, four of which were found to be spectroscopic binaries and will be published in a later paper. Also the two coldest targets (spectral type M4 or later) will be published in a later paper. This leaves thirteen M dwarfs and three late K dwarfs, see Table \ref{sampleinfo}. Spectra of all targets were obtained in the $J$-band (1100-1400 nm) with the CRIRES spectrograph at ESO-VLT \citep{Kaeufl2004}. The observations were carried out in service mode during ESO period 90 (October 1$^{\rm st}$ 2012 to March 31$^{\rm st}$ 2013). For all targets a slit-width of 0$\farcs$4~was used, giving a resolving power of R $\sim$ 50,000. Each target was observed in five wavelength intervals centered at $\lambda_C$~=~1177.9, 1188.0, 1205.2, 1258.4, 1303.2~nm. For each target a telluric standard star at similar air-mass was observed. These were mainly B-type stars that have a featureless spectra in the wavelength regions we are observing. 

Table \ref{sampleinfo} gives the signal-to-noise ratio (S/N) per pixel measured at the continuum level for the final spectrum of each target. Eleven wavelength regions in the four spectra centered at $\lambda_C$~=~1177.9, 1188.0~nm were chosen for the calculation, where we through visual inspection of each spectrum ensured the wavelength regions have no telluric lines and minimal influence of FeH lines. The observations had been planned in such a way as to achieve a S/N of $\sim$ 140 for stars with $J\le7.5$ and $\sim$80 for fainter stars, adjusting the exposure times according to $J$ magnitude. For telluric standard stars, a S/N of 600 was adopted. However, when observing with the CRIRES instrument one is required to choose among a set of fixed integration times. The actual exposure times and thus S/N values deviated therefore from the optimal ones by up to 20\%. The final S/N values for each target are furthermore affected by the division with the telluric standard star spectrum. These considerations may explain the variation of S/N values seen in Table \ref{sampleinfo}.

\section{Data processing and analysis}
\subsection{Data reduction}
\label{sect:datareduction}
The observational data was reduced with the CRIRES pipeline, version 2.3.2. Observed frames were dark- and flat-field corrected and an initial wavelength calibration based on thorium and argon lines was applied. Data from the two nodding positions were added into one spectrum per target. 

In our analysis we only used data obtained with the two central ones of the four chips making up the CRIRES detector (hereafter chip 2 and chip 3), since the two outer ones at the time of the observations were vignetted and contaminated by overlapping orders. In the pipelined spectra we found two unphysical very strong absorption spikes present in all spectra observed with chip 2. The features affected only a limited wavelength region, each less than 1~\AA. We chose to set the flux values in those regions to zero to avoid affecting the determination of stellar parameters. Furthermore, in the spectra of some of the telluric standard stars a very wide ($\sim$22~\AA) emission feature was present. Because we need these spectra to remove the telluric lines from the science spectra we chose to set the flux values in this wavelength region, of affected program stars, to zero.

\begin{table*}
\caption{Photometry for both the sample in this paper and L16.} 
\label{photometry}
\centering
\begin{tabular}{lccccccc}
\hline\hline
Target 			& $V$ 					& $J$ 			& $H$ 			& $K\rm_s$ 		& Parallax  [mas] 		& Ref.	\\ \hline
GJ~176			& 10.007 [0.04]				& 6.462 [0.024]		& 5.824 [0.033] 	& 5.607 [0.034] 	& 106.32 [0.60]			& 1, 5		\\
GJ~179			& 11.963 [0.04]				& 7.814 [0.024]		& 7.209 [0.046]		& 6.942 [0.018]		& 80.71 [0.33]			& 1, 5 		\\
GJ~203			& 12.448 [0.01]				& 8.311 [0.021]		& 7.840 [0.033]		& 7.542 [0.017]		& 113.5 [5.01]			& 1, 6 		\\
GJ~228 			& 10.420 [0.03]				& 6.795 [0.024]		& 6.306 [0.040]		& 6.032 [0.023]		& 91.65 [3.50]			& 1, 6 		\\
GJ~250B			& 10.045 [0.016]			& 6.579 [0.03]		& 5.976 [0.055]		& 5.723 [0.036] 	& 114.81 [0.44]$^b$		& 2, 6		\\
GJ~334			& 9.637 [0.01]				& 6.641 [0.018]		& 5.982 [0.040]		& 5.757 [0.017] 	& 68.79 [0.28]			& 1, 5 		\\
GJ~317			& 11.975 [0.04]				& 7.934 [0.027]		& 7.321 [0.071] 	& 7.028 [0.020] 	& 65.3 [0.4]			& 1, 7 		\\
GJ~433			& 9.821 [0.05]				& 6.471 [0.018] 	& 5.856 [0.036]		& 5.623 [0.021] 	& 109.57 [0.38]			& 1, 5 		\\
GJ~436			& 10.613 [0.01]				& 6.900 [0.024]		& 6.319 [0.023]		& 6.073 [0.016]		& 102.58 [0.31]			& 1, 6 		\\
GJ~514			& 9.029 [0.00]$^a$			& 5.902 [0.018]		& 5.300 [0.033]		& 5.036 [0.017]		& 130.62 [1.05]			& 3, 6 		\\
GJ~581			& 10.560 [0.02]				& 6.706 [0.026]		& 6.095 [0.033]		& 5.837 [0.023]		& 158.64 [0.35]			& 1, 5 		\\
GJ~628			& 10.101 [0.03]				& 5.950 [0.024]		& 5.373 [0.040]		& 5.075 [0.024]		& 232.29 [0.49] 		& 1, 5 		\\
GJ~674			& 9.407 [0.01]$^a$			& 5.711 [0.019]		& 5.154 [0.033]		& 4.855 [0.018]		& 220.24 [1.42]			& 3, 6  		\\
GJ~825			& 6.690 [0.00]			& 4.046 [0.266]		& 3.256 [0.216]		& 3.100 [0.230] 	&  251.14 [0.57]		& 1, 5 		\\
GJ~832			& 8.485 [0.08]				& 5.349 [0.032]		& 4.766 [0.256]		& 4.501 [0.018] 	& 201.12 [0.51]			& 1, 5 		\\
GJ~849			& 10.365 [0.03]				& 6.510 [0.024]		& 5.90 [0.04]		& 5.594 [0.017]		& 113.25 [0.32]			& 1, 5 		\\
GJ~872B			& 12.36 [0.25] 				& 7.944 [0.021]		& 7.473 [0.021]		& 7.300 [0.018]		& 61.36 [0.19]$^b$		& 4, 6 		\\
GJ~876			& 10.195 [0.03]				& 5.934 [0.019]		& 5.349 [0.049]		& 5.010 [0.021]		& 214.05 [0.97]			& 1, 5 		\\
GJ~880			& 8.638 [0.01]$^a$			& 5.360 [0.020]		& 4.800 [0.036] 	& 4.523 [0.016]		& 145.98 [0.51] 		& 3, 5 		\\
GJ~908			& 9.416 [0.03]				& 5.827 [0.023]		& 5.282 [0.031]		& 5.043 [0.020] 	& 169.09 [0.38]			& 1, 5 		\\
GJ~3634			& 11.926 [0.02]				& 8.361 [0.023]		& 7.76 [0.05]		& 7.470 [0.027]		& 50.55 [1.55]			& 1, 8 		\\
GJ~9356			& 9.878 [0.13]				& 7.660 [0.026]		& 7.074 [0.036]		& 6.963 [0.029]		& 32.93 [1.38]			& 1, 6 		\\
GJ~9415			& 12.237 [0.04]   			& 8.217 [0.029]		& 7.703 [0.044]		& 7.413 [0.021]		& 128.52 [3.90]			& 1, 6 		\\
HIP~12961		& 10.215 [0.06]				& 7.558 [0.021]		& 6.927 [0.031]		& 6.736 [0.018]		& 42.97 [0.31]			& 1, 5 		\\
HIP~31878		& 9.881 [0.08]			& 7.301 [0.020]		& 6.643 [0.024]		& 6.500 [0.020] 	& 45.33 [0.24]			& 1, 5 		\\
HIP~57172B		& 10.491 [0.01]				& 7.940 [0.026] 	& 7.29 [0.05] 		& 7.107 [0.024] 	& 33.51 [0.28]			& 1, 5 		\\
\hline
\end{tabular}
\noindent
\begin{flushleft}
{\bf References.} (1) \citet{Zacharias2013}, (2) \citet{Mann2015}, (3)~\citet{Koen2010}, (4)~\citet{Salim2003}, (5) \emph{Gaia DR1}, \citep{Gaia2016a, Gaia2016b}, (6) \emph{Hipparcos catalogue} \citep{vanLeeuwe2007},  (7) \citet{Anglada-Escude2012}, (8)~\citet{Riedel2010}.\\
\noindent
\newline
{\bf Notes.} $V$ magnitudes are given in the Johnson system. J, H, and $K_{\rm s}$ magnitudes were taken from the 2MASS catalogue \citep{Cutri2003}. \\
$^a$ Photometric errors are for the absolute magnitude since \citet{Koen2010} did not include any errors for the apparent magnitude. \\
$^b$ Parallax values are for the brighter binary companion.
\end{flushleft}
\end{table*}

In addition to the data reduction by the pipeline we performed several post-processing reduction steps to improve the quality of the spectra. We applied a continuum rectification and refined wavelength calibration to all science and telluric standard spectra. Accurate determination of the stellar parameters requires a reliable continuum placement. The continuum level of each spectrum was defined by selection of several points in regions that we identified as continuum, regions without any visible stellar or telluric lines, and each spectrum was individually rectified using a cubic spline interpolation. In the infrared the number of thorium and argon lines is not sufficient to ensure the precise wavelength correction needed, hence a refined wavelength calibration was done using telluric lines in a high-resolution solar spectrum \citep{Livingston1991}. In the spectra from chip 2 for $\lambda_C$~=~1258.4~\AA, only four telluric lines are present, and one of them cannot be used because it is overlapping with one of the unphysical absorption spikes discussed above. As a result no additional wavelength correction could be made, and these spectra were not used in the analysis. Lastly, we noticed that the noise in the spectra observed with chip 2 for $\lambda_C$~=~1303.2~\AA, was higher than in the remaining observed spectra. To avoid possible inconsistencies due to this we chose to exclude these observations. This left eight spectral regions per target, each covering about 60~\AA. The spectra for each target and corresponding telluric standard were adjusted for the radial velocity of the science target found in the literature. Furthermore, we used SME with a synthetic spectrum calculated for $T_{\rm eff}$~=~3500~K and log~$g$~=~5.0 to find the velocity correction needed to bring all spectra onto the laboratory wavelength frame. Subsequently, the telluric lines were removed from the science spectra using the spectra of the telluric standards, an example is shown in Fig. \ref{example}. For some stars we have several observations (GJ~334, GJ~433, GJ~514, GJ~908, GJ~3634 and HIP~31878), for which spectra were co-added using an unweighted mean.

\subsection{Determination of stellar parameters}
The first model atmospheres for M dwarfs were developed in the 1960-70s (e.g. \citealt{Tsuji1966a, Tsuji1996b, Auman1969, Mould1976}). Today the most widely used ones are MARCS, PHOENIX and ATLAS \citep{Gustafsson2008, Hauschildt1999, Castelli2004}. The different atmospheric models use somewhat different assumptions, e.g. concerning convection, and input data, e.g. continuous opacities. Therefore the results might change slightly depending on the model grid used, but exploring this effect is outside the scope of this paper. As discussed earlier the low surface temperatures allow molecules to be formed, and in the range of effective temperatures typical for M dwarfs (2700 $<$ $T_{\rm eff}$ $<$ 4000~K), molecules are as important as atomic species. Dominating molecules are TiO and H$_2$O, but also other oxides and hydrides are important. Great improvements in the molecular line data have been made in the last decades, but these are still one of the major limiting factors working with cool dwarfs. 

For determination of stellar parameter we used the software package Spectroscopy Made Easy, SME\footnote{\url{http://www.stsci.edu/~valenti/sme.html}} \citep{Valenti1996, Piskunov2017}, version 522. SME computes synthetic spectra on the fly, based on a grid of model atmospheres and a list of atomic and molecular data provided by the user. SME includes MARCS and ATLAS model atmospheres, with a sub-grid of MARCS 2012 models for cool dwarfs. This sub-grid covers effective temperature between 2500-3900~K in steps of 100~K and surface gravity between 3 and 5.5 in steps of 0.5. For the majority of the targets in this paper we used this sub-grid, however, for GJ~334, GJ~9356, HIP~12961 and HIP~31878 we used the full grid of MARCS models (calculated in 2012 and distributed on the MARCS website\footnote{\url{http://marcs.astro.uu.se}}), which covers temperatures above the range of the sub-grid. 

Desired stellar parameters are determined through $\chi^2$ minimization between the synthetic and observed science spectrum. Our spectra do not contain sufficient information to determine the surface gravity, effective temperature and metallicity simultaneously. We therefore determined the surface gravity and effective temperature prior to the metallicity (see Sects.~\ref{sect:logg} and \ref{sect:teff}). We used the same approach in L16. Based on the good agreement found between the components of the binaries analyzed in L16 we consider the method to be reliable.

\subsubsection{Line data}
We used the same linelist as published in L16, where most of the atomic line data were taken from the VALD database\footnote{\url{http://vald.astro.uu.se}} \citep{Kupka2000, Heiter2008} with some additional data from \citet{Melendez1999}, and oscillator strengths and van der Waals broadening parameters were adjusted using a high-resolution solar spectrum. For more details see L16, section 4.3.1 and Table A.1 in that paper. We only used molecular data for one species, FeH, for which the data was provided by Bertrand Plez, available on the MARCS webpage. For the analysis we used the solar abundances by \citet{Grevesse2007}.

\subsubsection{Surface gravity}
\label{sect:logg}
To calculate the surface gravity we used the fundamental relation $g$~=~GM/R$^2$, where M is the stellar mass, R the stellar radius, and G the Newtonian constant of gravitation. In the recent years two independent empirical relations have been derived between photometry and the stellar mass for low-mass stars: \citet[hereafter M15]{Mann2015} and \citet[hereafter B16]{Benedict2016}. M15 provide a fourth-order polynomial fit between the absolute $K\rm_S$ magnitude and masses determined by stellar models, their Eq. 10, valid between $M_{K\rm_S}$~$\epsilon$~[4.6, 9.8]. They showed that their determined masses and radii were consistent with those determined from low-mass eclipsing binaries within $\pm$5\% for masses between 0.1-0.65~\Ms. B16 based their mass-luminosity relation on 47 stars in binaries with determined dynamical masses in a range from 0.08-0.6~\Ms. Two relations were determined using fifth-order polynomials to the $M_{K\rm_S}$ and $M_V$ magnitude, their Eq.~11, valid for $M_{K\rm_S}$~$<$~10 and $M_{V}$~$<$~19. We chose to use the $K\rm_S$ magnitude because of the reported lower rms value of the residuals to the polynomial fit and furthermore since most of our targets have lower observational uncertainty in $K\rm_S$-band compared to the $V$-band. Using the data in Table \ref{photometry} we calculated masses using both the relation by M15 and B16. The resulting differences were not systematic, and the resulting surface gravities differ by up to 0.07~dex. Because of the possible uncertainty in stellar models we chose to use the calibration by B16. 

M dwarfs are intrinsically faint hence the number for which a radius can be determined from interferometry is limited. Instead we used an empirical relation. The most recent and extensive was provided by M15. Four calibrations were determined using photometry or effective temperature, and with or without including the metallicity as a second independent variable. We used their Eq. 4 which gives the radius as a function of $K\rm_S$ magnitude, without dependance on metallicity. We chose to use the photometry over the effective temperature because of the lower uncertainties for individual targets. The inclusion of the metallicity only reduced the scatter of the polynomial fit marginally (see Sect. 6.1 of M15), hence it is safe to use only the photometry when deriving the radius for our targets. Table \ref{logg} shows the calculated values for the radius from the M15 relation, the masses from the B16 relation, and the corresponding base-10 logarithm of the surface gravity (log~$g$).

\begin{table}
\caption{Determined mass and surface gravity.} 
\label{logg}
\centering
\begin{tabular}{lccc}
\hline\hline
Target 			&  Mass [\Ms]		& Radius [\rsun]	& log~$g$ [cm s$^{-2}$]   \\ \hline
GJ~179			& 0.404			& 0.376			& 4.89 		\\
GJ~203			& 0.192			& 0.226			& 5.01 		\\
GJ~228 			& 0.524		 	& 0.468			& 4.82 		\\
GJ~334			& 0.638			& 0.622			& 4.66 		\\
GJ~433			& 0.527		 	& 0.472			& 4.81 		\\
GJ~514			& 0.562		 	& 0.505			& 4.78 		\\
GJ~825			& 0.627		 	& 0.593			& 4.69 		\\
GJ~832			& 0.491		 	& 0.442			& 4.84		\\
GJ~872B			& 0.449		 	& 0.409			& 4.87 		\\
GJ~880			& 0.601		 	& 0.550			& 4.74 		\\
GJ~908			& 0.460		 	& 0.417			& 4.86 		\\
GJ~3634			& 0.497			& 0.446			& 4.84 		\\
GJ~9356			& 0.638			& 0.698			& 4.56 		\\
GJ~9415			& 0.178		 	& 0.214			& 5.03 		\\
HIP~12961		& 0.640			& 0.630			& 4.65		\\
HIP~31878		& 0.643			& 0.652			& 4.62 		\\
\hline
\end{tabular}
\begin{flushleft}
\noindent
{\bf Notes.} Masses were determined using the photometric relation by B16 and the stellar radii were determined by the photometric relation by M15. Used photometric data can be found in Table \ref{photometry}.
\end{flushleft}
\end{table}

\subsubsection{Effective temperature}
\label{sect:teff}
Over the years several methods have been proposed to estimate the effective temperature of M dwarfs. In our previous study L16 we showed that for M dwarfs with effective temperatures colder than approximately 3400~K, spectral fitting of FeH lines present in spectra with $\lambda_C$~=~1205~\AA~and 1258~\AA~can be used to estimate the effective temperature. At these temperatures we found no degeneracy in line strength between the effective temperature, and the metallicity or surface gravity. The determination was done by calculating a grid of synthetic spectra using SME, stepping 0.1~dex in metallicity from $-$0.5 to +0.5~dex, and by 25~K from 3000 to 3800~K in effective temperature. For each synthetic spectrum in the grid a $\chi^2$ value was calculated for the selected FeH lines ($\sim$50 lines). As expected the line depth decreases with increasing temperature, and at 3800~K the FeH lines are too shallow and almost insensitive to changes of the temperature and calculation of the $\chi^2$ values is no longer meaningful.

In L16 we had access to the FeH lines in the spectra observed with chip 2, $\lambda_C$~=~1258~\AA, which was not available for the targets in this paper due to problems with the observations (see Sect.~\ref{sect:datareduction}). We therefore re-analyzed all targets in our previous paper without this wavelength region and concluded that this changes the effective temperature by maximum $\pm$25~K in all cases.

\begin{figure*}
\centering
\includegraphics[width=0.33\textwidth]{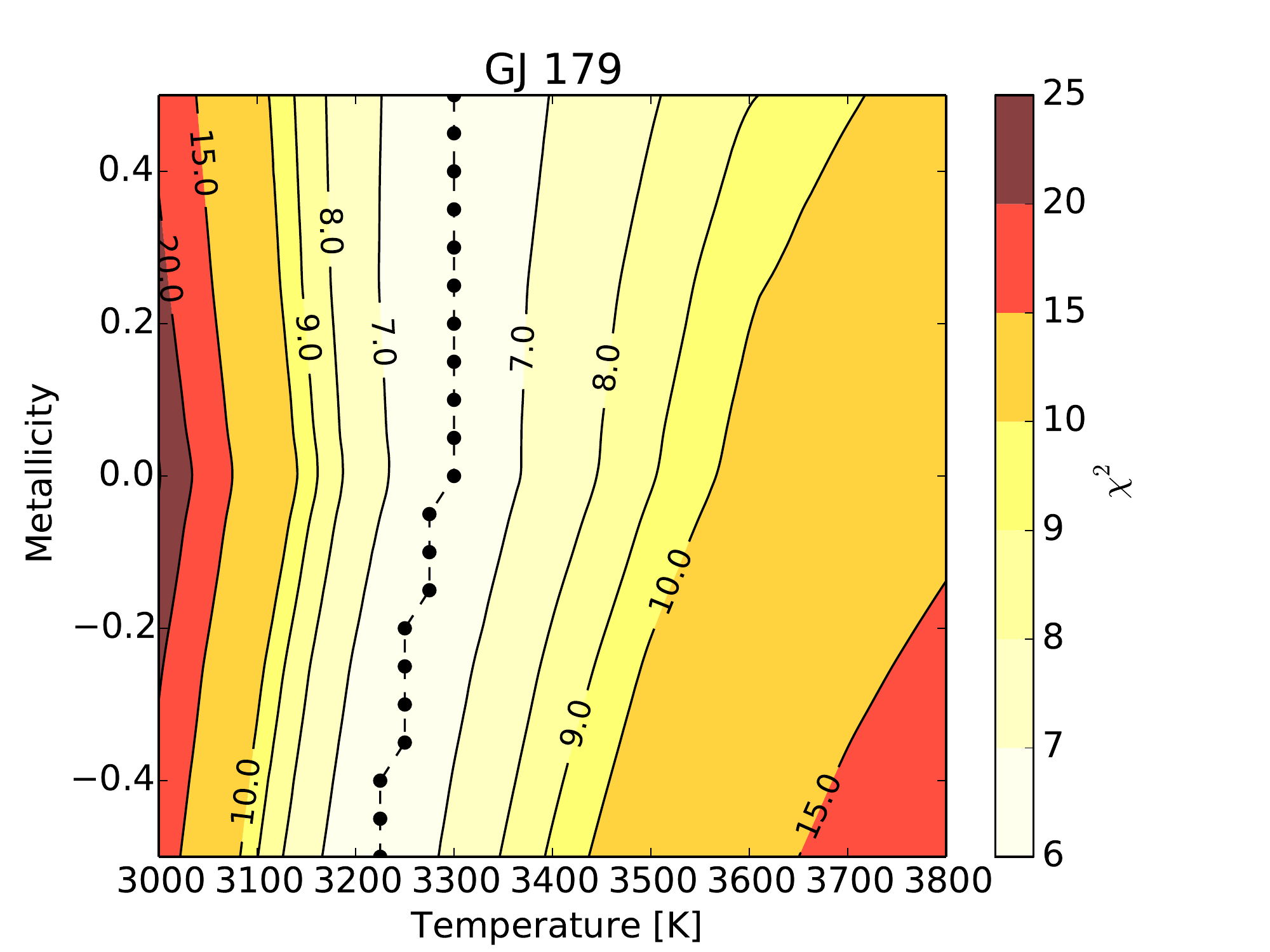} 
\includegraphics[width=0.33\textwidth]{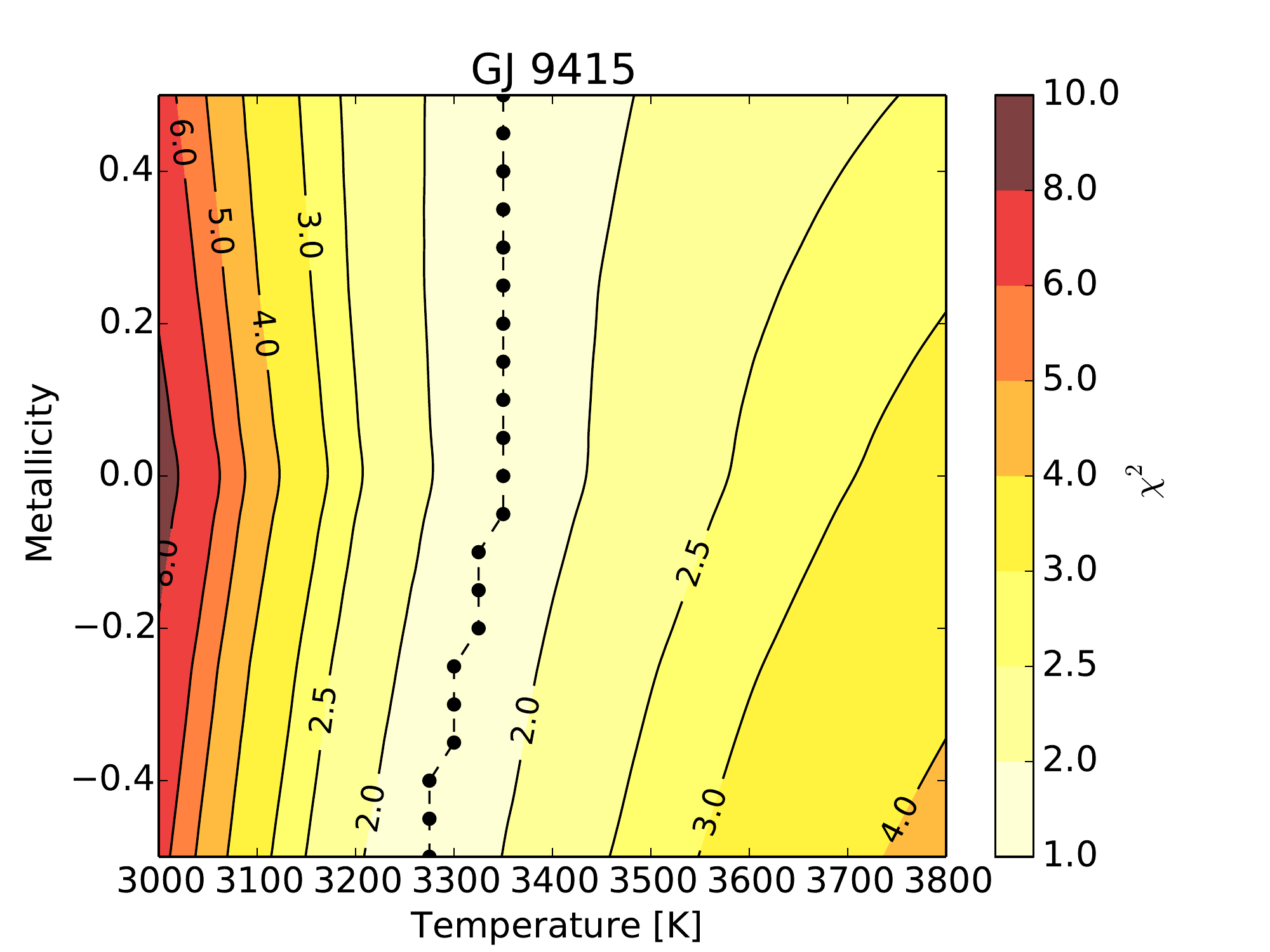}
\includegraphics[width=0.33\textwidth]{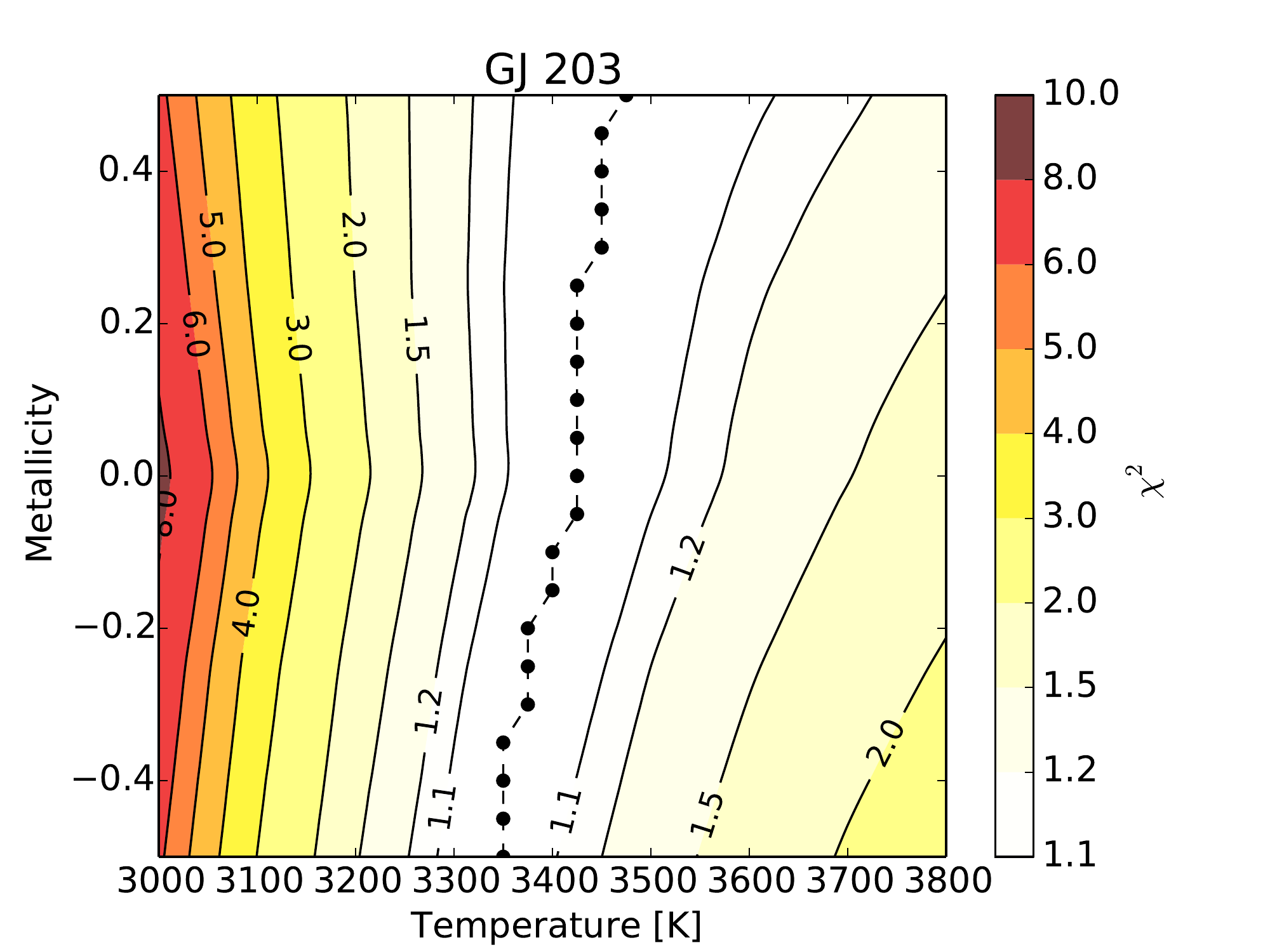}
\includegraphics[width=0.33\textwidth]{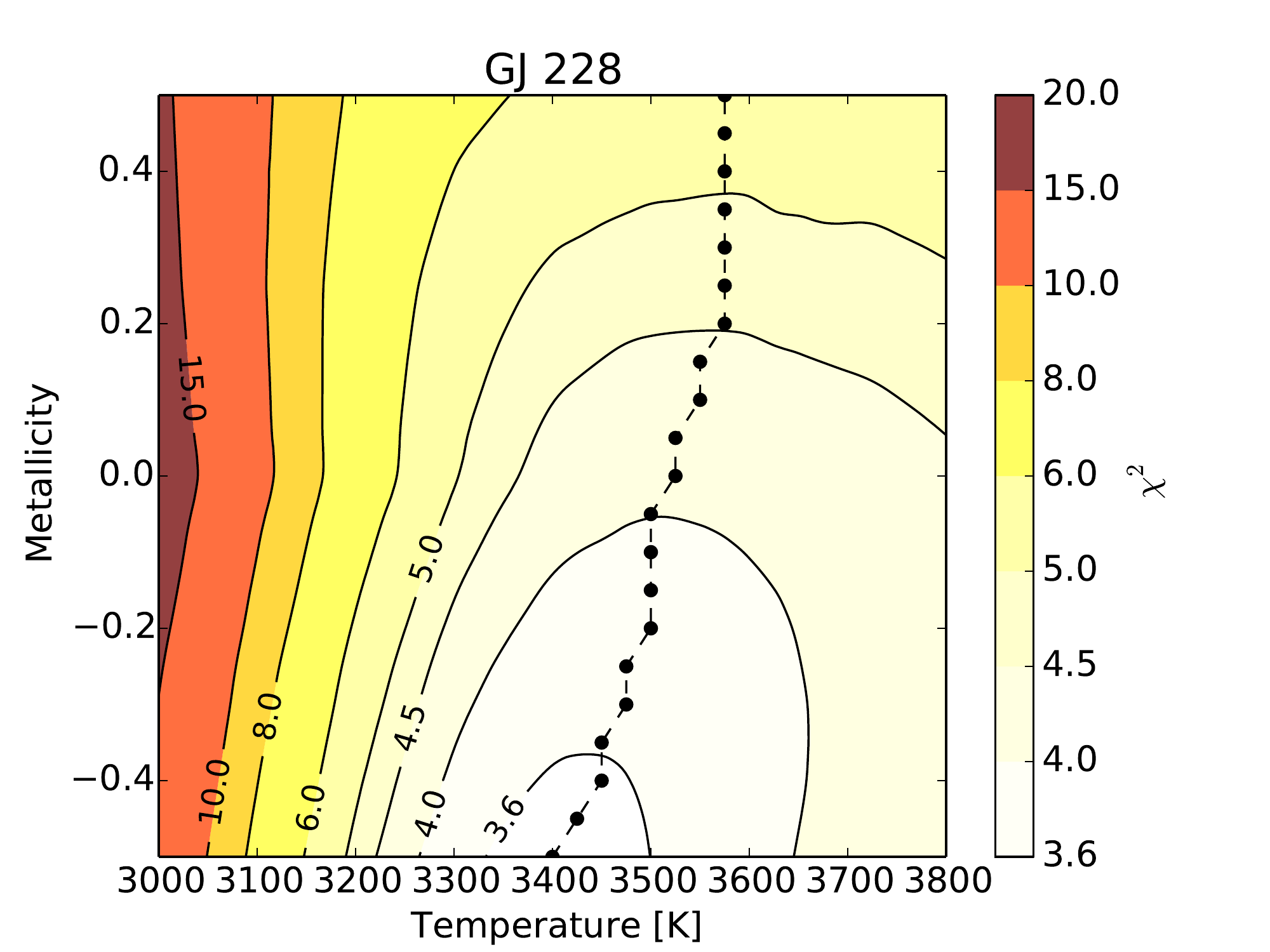}
\includegraphics[width=0.33\textwidth]{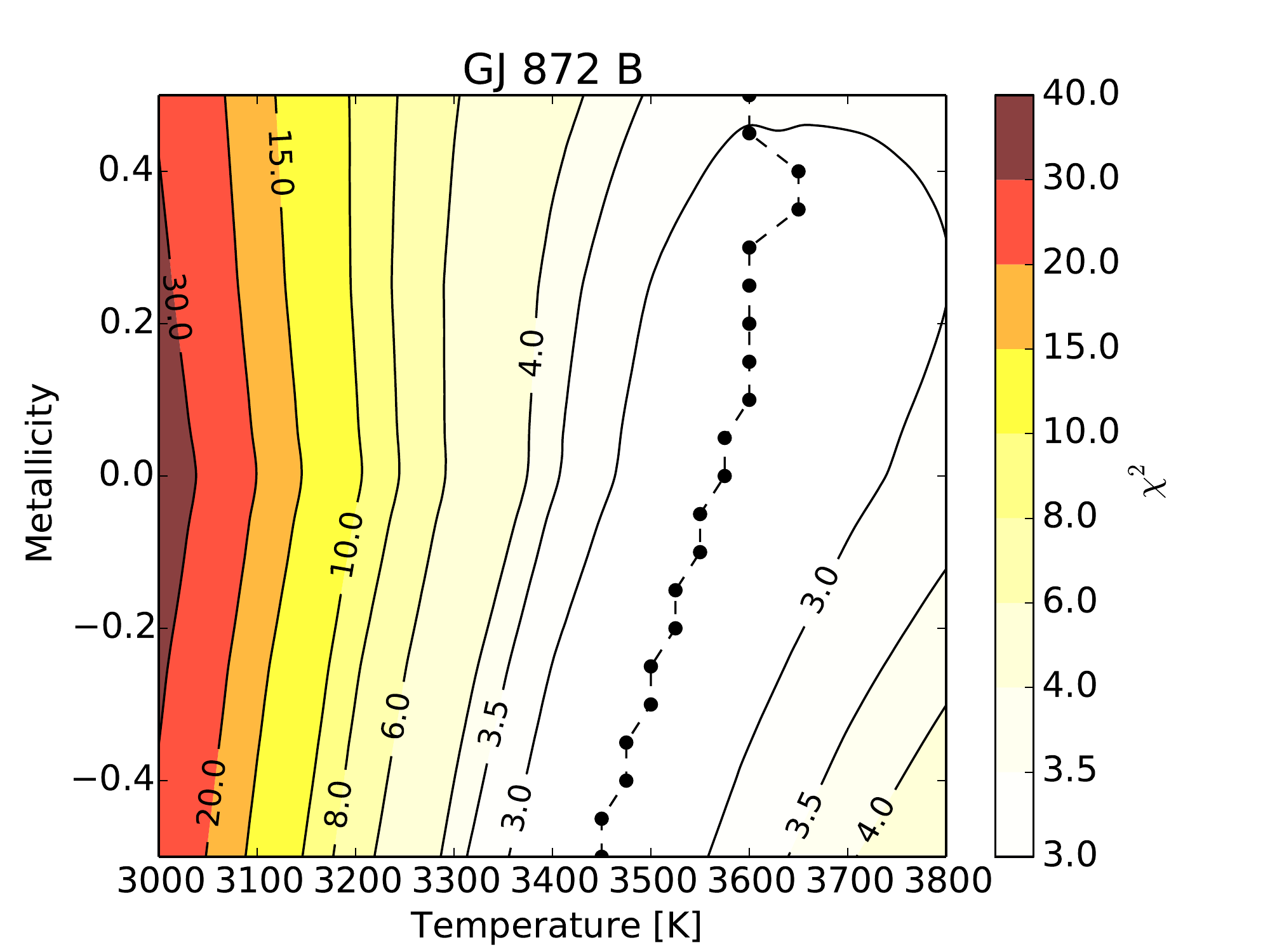}
\includegraphics[width=0.33\textwidth]{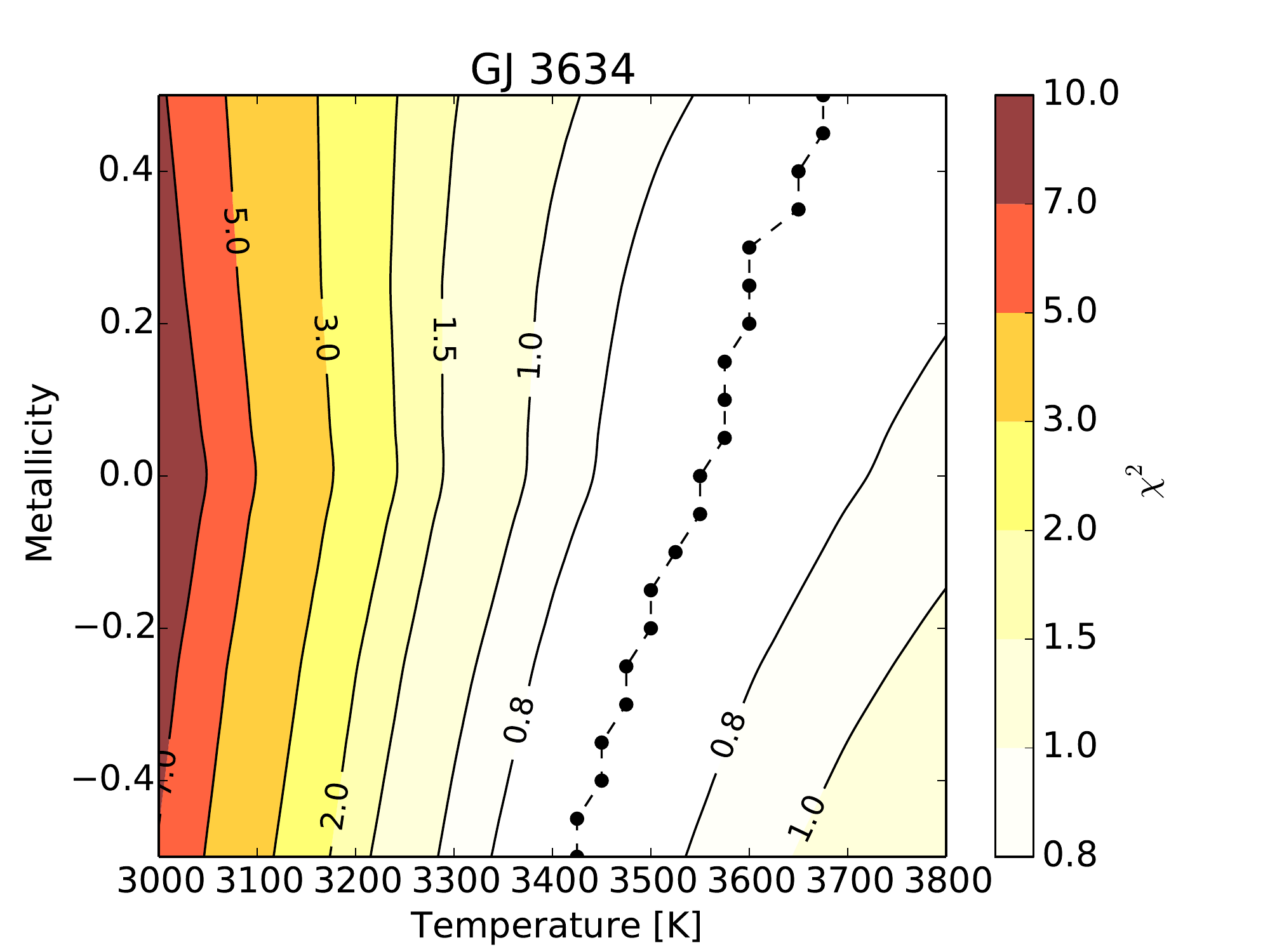}
\caption{Demonstration of FeH line strength dependency on effective temperature and metallicity for the six coldest targets in our sample. The contour plots in this figure show the calculated $\chi^2$ of the fit based on a grid of metallicity and effective temperature. The connected dots indicate the temperature with the minimum $\chi^2$ for each step in metallicity. The plots are shown in order of increasing degeneracy between metallicity and effective temperature. 
\label{FeH} }
\end{figure*}

We performed the analysis of the FeH lines for all targets with spectral types colder than M1. The majority of targets analyzed for this paper are however warmer than the sample in L16.  Shown in Fig.~\ref{FeH} is the result for the six coldest targets, and the warmer targets are shown in the Appendix. The dotted line shows the temperature corresponding to the minimum $\chi^2$ for each metallicity step. For targets GJ~179, GJ~203 and GJ~9415 we found that the temperature giving the lowest $\chi^2$ value is almost independent of the metallicity, while for GJ~228, GJ~872B and GJ~3634 we see a slight dependance on both. We decided to use the spectroscopic temperature if there was a maximum 200 degree temperature difference between found $\chi^2$ minimum for all metallicities. This value was chosen because we vary the temperature with $\pm$~100~K in our error analysis. Based on this criterion we used the spectroscopic temperature from the FeH analysis for GJ~179, GJ~203, GJ~228, GJ~872B and GJ~9415. 

The given effective temperatures in column 1 in Table~\ref{teff} correspond to the temperature at minimum $\chi^2$ at [M/H]~=~0.00. We compared our result with calculated values from two empirical photometric calibrations (\citealt{Casagrande2008}, M15) and spectroscopically determined values by \citet[hereafter RA12]{Rojas-Ayala2012} and \citet{Lepine2013}. The values from \citet{Casagrande2008} shown in Table \ref{teff} were calculated as an unweighted mean of temperatures calculated with the $V-J$, $V-H$ and $V-K$ colors. M15 found that their temperature relations show significant dependence on the metallicity. The effective temperatures in Table \ref{teff} were therefore calculated using the $V-J$ color from Table \ref{photometry} and metallicity determined by SME using the $T_{\rm eff}$ given by the FeH line strength. Both RA12 and \citet{Lepine2013} estimated the effective temperatures spectroscopically using Phoenix atmospheric models. RA12 used their so-called H$_2$O-K2 index, calibrated against BT-Settl-2010 models \citep{Allard2010}. The index is given by the flux difference in specific wavelength segments in the $K$-band that should trace the change of the spectral shape due to water absorption and thereby the temperature. \citet{Lepine2013} instead estimated the temperature by comparing their observed low- and medium resolution optical spectra to a grid of pre-calculated synthetic BT-Settl spectra \citep{Allard2011} at different temperatures, surface gravities and metallicities. 

\begin{table}
\caption{Estimated effective temperatures (in K) from this work compared to four different studies.}  
\label{teff}
\centering
\begin{tabular}{lccccccc}
\hline\hline
Target  		& Our FeH plots 	& C08		& M15		& RA12 		& L13  	\\
                		& (1)            		& (2)   		& (3)           	& (4)  		& (5) 	\\ \hline
GJ~179		& 3300			& 3095		&  3383		& 3424  		& 3260	\\
GJ~203		& 3425			& 3128		&  3297		&			& 3290	\\
GJ~228		& 3525			& 3342		&  3457		&			& 3550	\\
GJ~872B		& 3575			& 3045		&  3192		& 3569		&		\\
GJ~3634		& 3550			& 3332		&  3538		&			& 		\\
GJ~9415		& 3350			& 3163		&  3359		&			&		\\
\hline
\end{tabular}
\begin{flushleft}
\noindent
\newline
{\bf Notes.} Column 1 contains the temperatures estimated in this work using spectral fitting of FeH lines. Column 2 contains values calculated with the photometric calibration by C08 \citep{Casagrande2008}, and Column 3 values calculated with the photometric calibration by M15. Columns 4 and 5 show the adopted values in the spectroscopic works by RA12 and L13 \citep{Lepine2013}.
\end{flushleft}
\end{table}

Our values using the FeH line strength are consistently around 300~K higher than those by \citet{Casagrande2008}. This was expected since their calibration builds on the assumption that the star can be approximated as a blackbody beyond 2 $\mu$m, but for M dwarfs a large amount of the flux is shifted towards the infrared \citep{Rajpurohit2013}. Compared to the remaining studies we agree within 150~K. The exception is GJ~872B, where the value for M15 is substantially lower. However, we expect this is due to problems with the photometry for this star, which will be discussed in more detail in Sect.~\ref{sect:CMD}. As an additional indication that this is due to the photometry the spectroscopic method by RA12 gives a very similar temperature to ours.

For the other warmer targets we used three different ways to determined the effective temperature. For GJ~514, GJ~880 and GJ~908 there is a determination of the effective temperature by M15 based on a comparison with optical spectra and CFIST suite of the BT-SETTL version of the PHEONIX model atmosphere models, with a stated typical error of 60~K. For the remaining targets we used the photometric calibration by M15. Because of the dependence found on metallicity we determined the temperature in an iterative process until the change in determined temperature was less than 50~K. For five targets we found discrepancies in temperature sensitive lines when we inspected the calculated synthetic spectra with SME. For these targets we determined the effective temperature using SME, see Sect.~\ref{sect:metallicity} for more details. Adopted effective temperatures for all targets can be found in Table~\ref{metallicity}.

\subsubsection{Metallicity}
\label{sect:metallicity}
All metallicities derived in this work refer to the overall metallicity [M/H] determined by the fit of lines from all atomic species available in our spectra. The available wavelength regions result in about 20 useful lines, from species Fe, Ti, Mg, Ca, Si, Cr, Co,  Mg. We determined the metallicity by letting SME simultaneously vary the metallicity and macroturbulence ($\zeta_t$), while all other parameters where kept fixed. Five targets had previous estimates for the projected rotational velocity $v$~sin$i$, see Table \ref{metallicity}. For the other targets we tested $v$~sin$i$~=~1.0, 2.0 and 3.0~\kms~and found that the metallicity determinations varied by less than 0.05~dex, while the change in macroturbulence was more significant. Using additional tests allowing the metallicity and $v$~sin$i$, and secondly the metallicity, $v$~sin$i$ and $\zeta_t$, to vary simultaneously we concluded that there is a degeneracy between these two broadening parameters, but the metallicity remained relatively consistent independent of the combination used. For the analysis we used $v$~sin$i$~=~2.0~\kms~for targets resulting in a determined macroturbulence above one, and for the other targets we used $v$~sin$i$~=~1.0~\kms. We note that neither the determined values for $v$~sin$i$ or $\zeta_t$ should be seen as physical values, exception the five stars where $v$~sin$i$ was determined by independent methods, see Table \ref{metallicity}.

We found noticeable effect on the determined metallicity depending on the used microturbulence ($\xi_t$). As we only have access to 15-20 lines and many with similar strength we do not have enough spectral information to treat $\xi_t$ as a free parameter in SME. \citet{Wende2009} investigated the atmospheric velocities in M dwarfs using the 3D radiative hydrodynamical code CO$^5$BOLT. They found that the microturbulence decreases with lower temperature and higher surface gravity.

The difference depending on the surface gravity in the parameters range of our targets is negligible compared to the effect of the effective temperature. We estimated the microturbulence from their Fig.~10 that shows the micro- and macroturbulence as function of effective temperature,  where $\xi_t$ varies from 0.1 to 0.5~\kms~going from 2700 to 4000~K. For the warmer targets we relied on the simulations by \citet{Steffen2013} that investigated warmer stars also using the CO$^5$BOLT code. Their coldest simulation was for $T_{\rm eff}$~=~4500~K and log~$g$~=~4.5, and we used this for the four targets where we derive effective temperatures from 4000-4500~K. The reported value of 0.7~\kms~fit well with the trend seen for the M dwarfs in \citet{Wende2009}. Used values for all targets in the analysis for the metallicity determination are given in column 5 in Table \ref{metallicity}.

\begin{figure}
\centering
\includegraphics[width=0.4\textwidth]{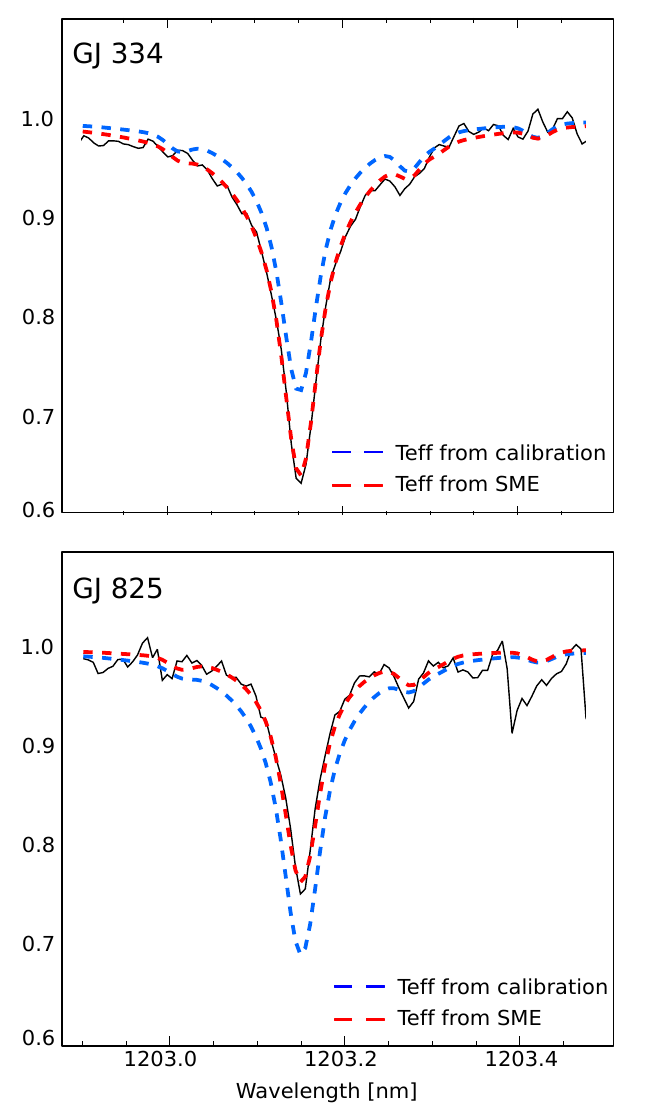} 
\caption{Wavelength interval around the Si line at 1203.15~nm~for targets GJ~334 and GJ~825. The observed spectrum is shown with the solid black line, and calculated synthetic spectra at two different temperatures at best-fit metallicity are shown as dashed lines. The blue spectrum is calculated at the temperature given by the empirical calibration by M15, while the red one results from letting SME simultaneously vary the effective temperature, metallicity and macroturbulence.
\label{Si} }
\end{figure}

\begin{table*}
\caption{Determined stellar parameters.} 
\label{metallicity}
\centering
\begin{tabular}{lcccccc}
\hline\hline
Target 			&  Metallicity 			&  $T_{\rm eff}$	 [K]	 	& $\zeta_t$ [\kms] 		& $\xi_t$ [\kms]			& $v$~sin$i$ [\kms]  &  Ref.  \\ \hline
GJ~179			& +0.36 $\pm$ 0.04		& 3300				& 0.06				& 0.25				& 1.00		\\
GJ~203			& $-$0.13	$\pm$ 0.04	& 3425				& 0.02				& 0.25				& 1.00		\\
GJ~228 			& $-$0.29 $\pm$ 0.04	& 3525				& 1.19				& 0.30				& 1.00		\\
GJ~334			& +0.13 $\pm$ 0.06 		& 4044				& 1.70				& 0.70				& 2.00		\\
GJ~433			& $-$0.02 $\pm$ 0.05	& 3616				& 0.30				& 0.35				& 1.00		\\
GJ~514			& +0.07 $\pm$ 0.07 		& 3727				& 0.27				& 0.35				& 1.30		& 1, 6 \\
GJ~825			& $-$0.01 $\pm$ 0.04 	& 3795				& 0.15				& 0.40				& 3.30		& 2 	\\
GJ~832			& $-$0.06 $\pm$ 0.04 	& 3707				& 0.29				& 0.35				& 1.00		\\
GJ~872B			& $-$0.21 $\pm$ 0.03 	& 3575				& 0.42				& 0.30				& 1.00		\\
GJ~880			& +0.20 $\pm$ 0.05 		& 3720				& 0.15				& 0.35				& 2.07		& 3, 6 \\
GJ~908			& $-$0.51 $\pm$ 0.05 	& 3646				& 3.70				& 0.35				& 2.25		& 4, 6 \\
GJ~3634			& +0.04 $\pm$ 0.06 		& 3536				& 0.19				& 0.30				& 1.00 		\\
GJ~9356			& +0.16 $\pm$ 0.04		& 4526				& 0.10				& 0.70				& 1.00		\\
GJ~9415			& $-$0.38 $\pm$ 0.04 	& 3350				& 3.39				& 0.25				& 2.00 		\\
HIP~12961		& +0.10 $\pm$ 0.06 		& 4131				& 2.55				& 0.70				& 1.50		& 5	\\
HIP~31878		& $-$0.04 $\pm$ 0.05 	& 4246				& 5.69				& 0.70				& 2.00		\\
\hline
\end{tabular}
\begin{flushleft}
\noindent
{\bf References}
$v$~sin$i$: (1)~\citet{Browning2010}, (2)~\citet{Houdebine2010}, (3)~\citet{Torres2006}, (4)~\citet{Bonfils2011}, (5)~\citet{Forveille2011}. Effective temperature: (6)~\citet{Mann2015}. \\
\noindent
\newline
{\bf Notes.} Metallicity and macroturbulence determined in this work using SME. The values used for the surface gravities are given in Table~\ref{logg}. Given uncertainties for the metallicity are the co-added differences changing the effective temperature by $\pm$100~K and surface gravity by $\pm$0.1~dex. \\

We note that except for the stars with previous determination of $v$~sin$i$ neither the given values for $\zeta_t$ or $v$~sin$i$ should be seen as physical values because of the degeneracy between those broadening parameters. We give those values for reproducibility of our results.
\end{flushleft}
\end{table*}

As discussed in the previous section we found a rather poor fit in the calculated synthetic spectrum for the silicon lines for five targets when using the temperature by the calibration by M15. These lines become significantly stronger and develop wings at higher temperatures. For GJ~334, GJ~9356, HIP~12961 and HIP~31878 all silicon lines were too weak compared to the observed spectrum, hence we concluded that the temperatures from the calibration by M15 were too low. The decreased accuracy for these warmer targets is not surprising since they are outside or on the limit for the calibration by M15 of $T_{\rm eff}$~$\epsilon$~[2700, 4100]. For GJ~825 we found the opposite where the silicon lines instead were too strong using the temperature by the M15 calibration. The reason for the discrepancy for this target is more unclear, but most likely due to the photometry where the $J$, $H$ and $K$ magnitudes have among the largest uncertainties of all targets in this paper. Because these spectra contain both temperature- and metallicity sensitive lines, we used SME to determine the effective temperature for these five targets by letting SME simultaneously solve for [M/H], $T_{\rm eff}$ and $\zeta_t$. The found best-fit temperatures were 100-250~K different and gave much better agreement between observed and synthetic spectra. An example from one of the Si line can be seen in Fig.~\ref{Si}. 

To estimate the uncertainty of the metallicity due to the uncertainties in the adopted effective temperature and surface gravity we varied $T_{\rm eff}$ by $\pm$100 K and log~$g$ by $\pm$0.10 dex. One parameter was varied at a time and new values for the metallicity were derived with SME. We furthermore included the effect of uncertainties due to the microturbulence. No estimates of the error in the derived theoretical values can be found in \citet{Wende2009} nor \citet{Steffen2013}, however both point out that the theoretical predictions from 3D models tend to be lower than determinations for solar-like stars and the attempts by \citet{Bean2006} for M dwarfs. Based on the difference between the values in \citet{Bean2006} using synthetic spectra in the optical compared to the values for the same effective temperature given by \citet{Wende2009} we chose to use 0.5~\kms~as an offset. For consistency we used the same value for all targets and calculated new metallicities with SME with this higher microturbulace. The differences between the original and the new metallicity and the four combinations for $T_{\rm eff}$ and log~$g$ were added in quadrature. The stellar parameters and the estimated uncertainties for the metallicity are given in Table~\ref{metallicity}. 

\begin{figure}
\centering
\includegraphics[width=0.5\textwidth]{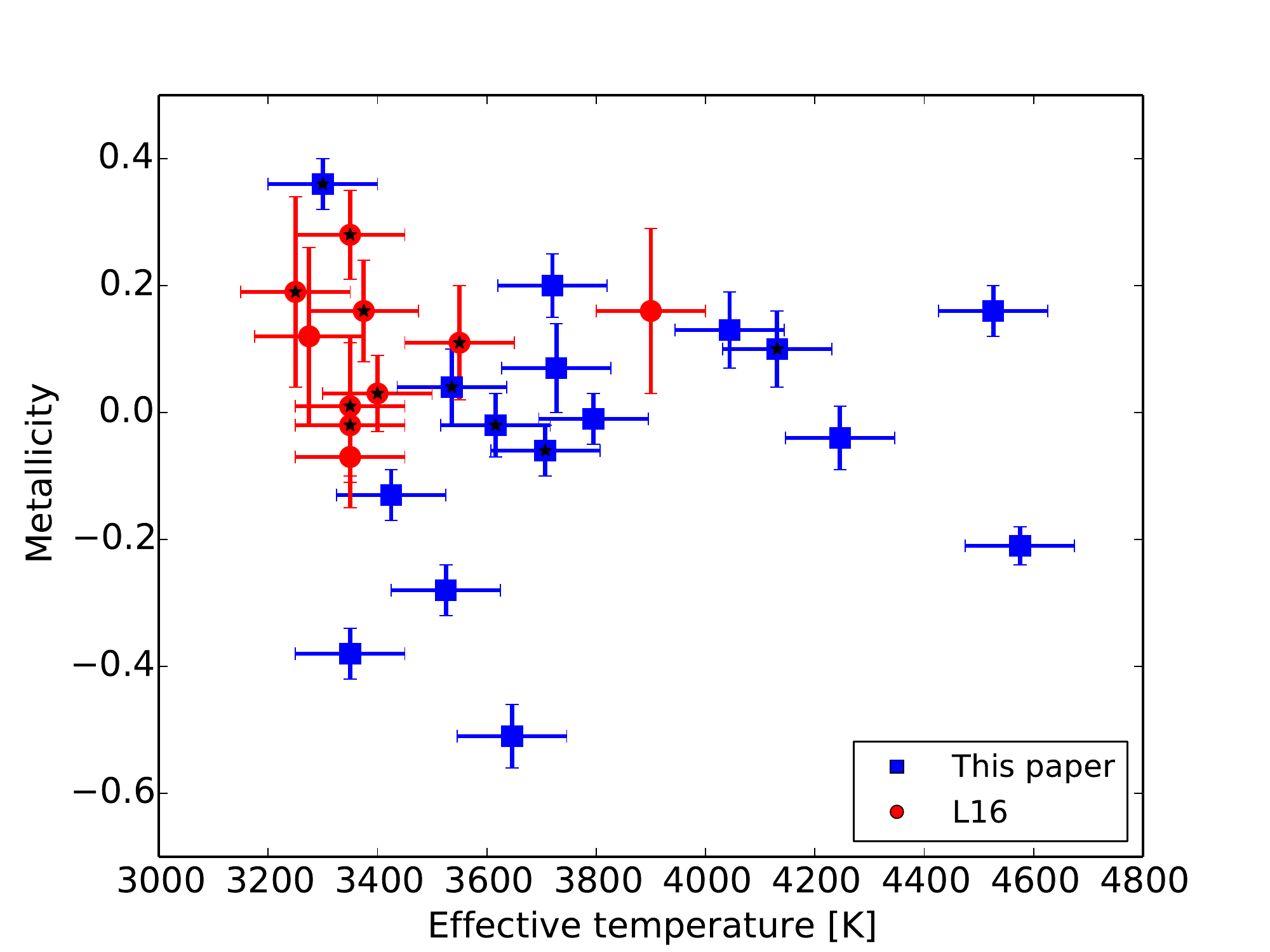} 
\caption{Distribution of the metallicity and effective temperature of our two samples. The blue squares show the results for the targets presented in this work (Table~\ref{metallicity}), and the red circles the results from L16. The uncertainties in effective temperature are $\pm$100~K for all targets. Data points marked with a small star symbol indicate targets with confirmed planet(s).
\label{meh_vs_teff} }
\end{figure}

\section{Results and discussion}
We have determined the metallicity, effective temperature and surface gravity for sixteen cool dwarf stars, ranging in spectral type from K5 to M3.5. In Fig.~\ref{meh_vs_teff} we show the distribution of the two samples in the metallicity $-$ effective temperature plane. Including both samples the analyzed stars spread approximately one~dex in metallicity and 1400~K in temperature.

\begin{figure}
\centering
\includegraphics[width=0.48\textwidth]{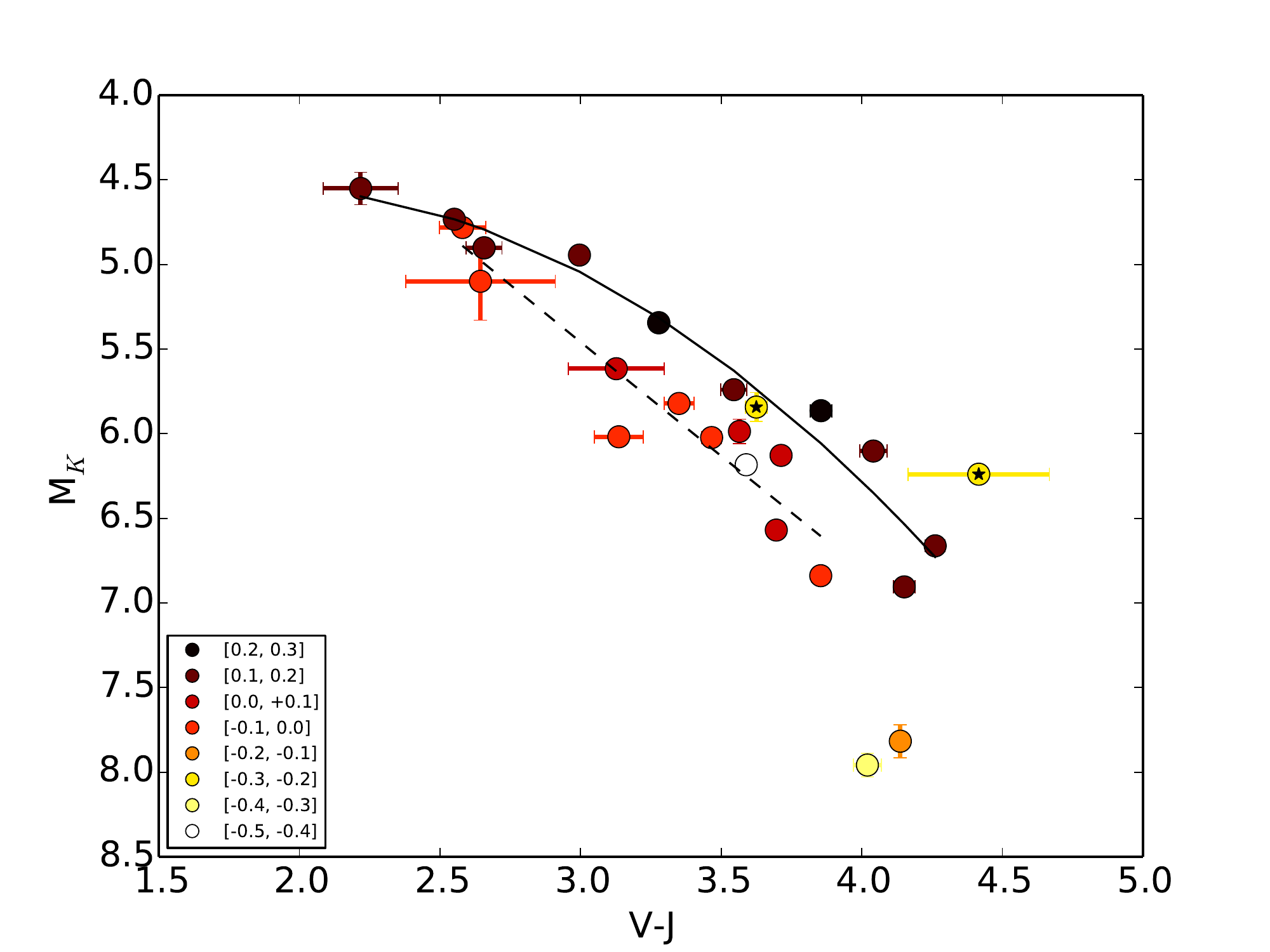} 
\includegraphics[width=0.48\textwidth]{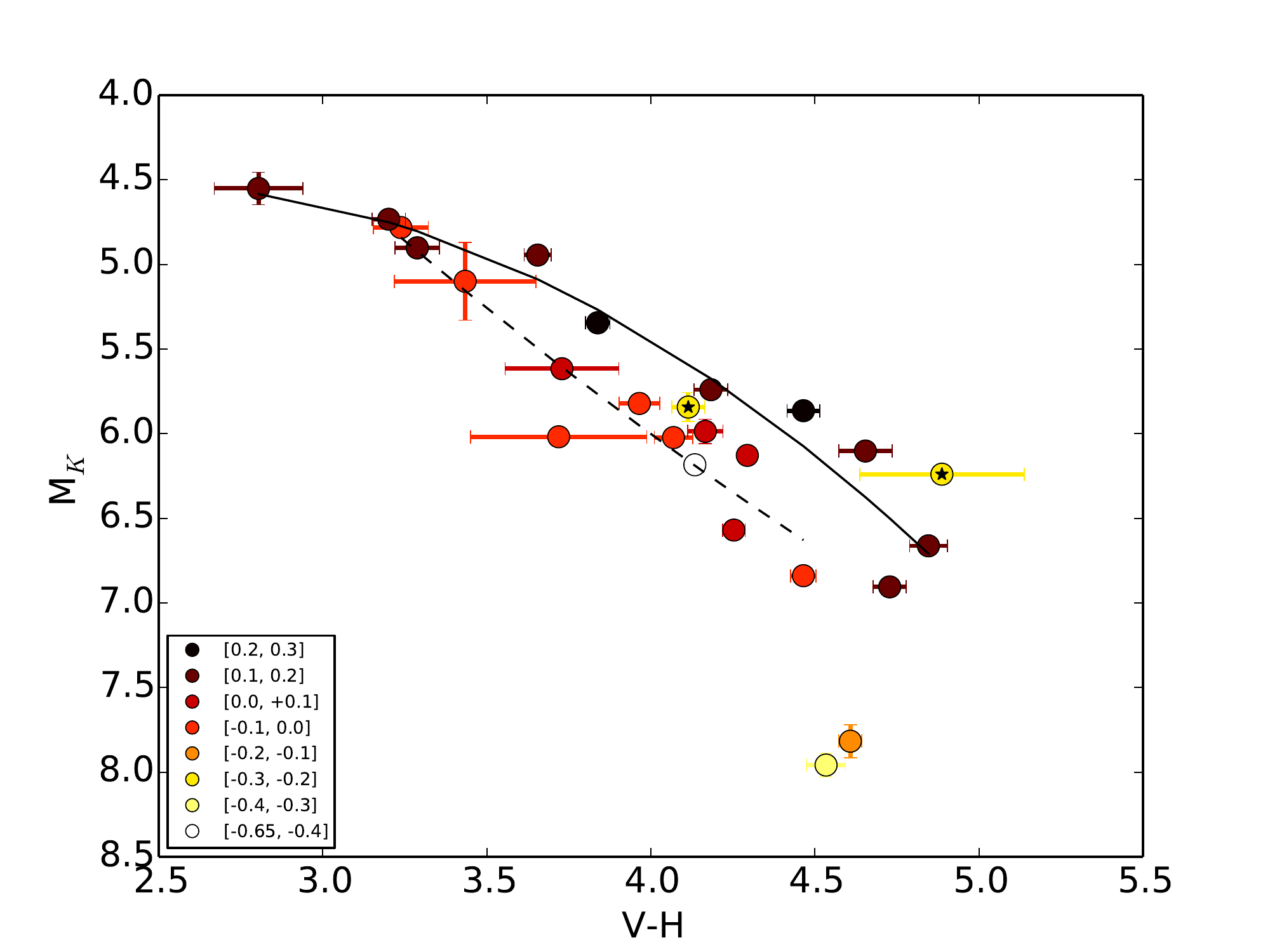}
\includegraphics[width=0.48\textwidth]{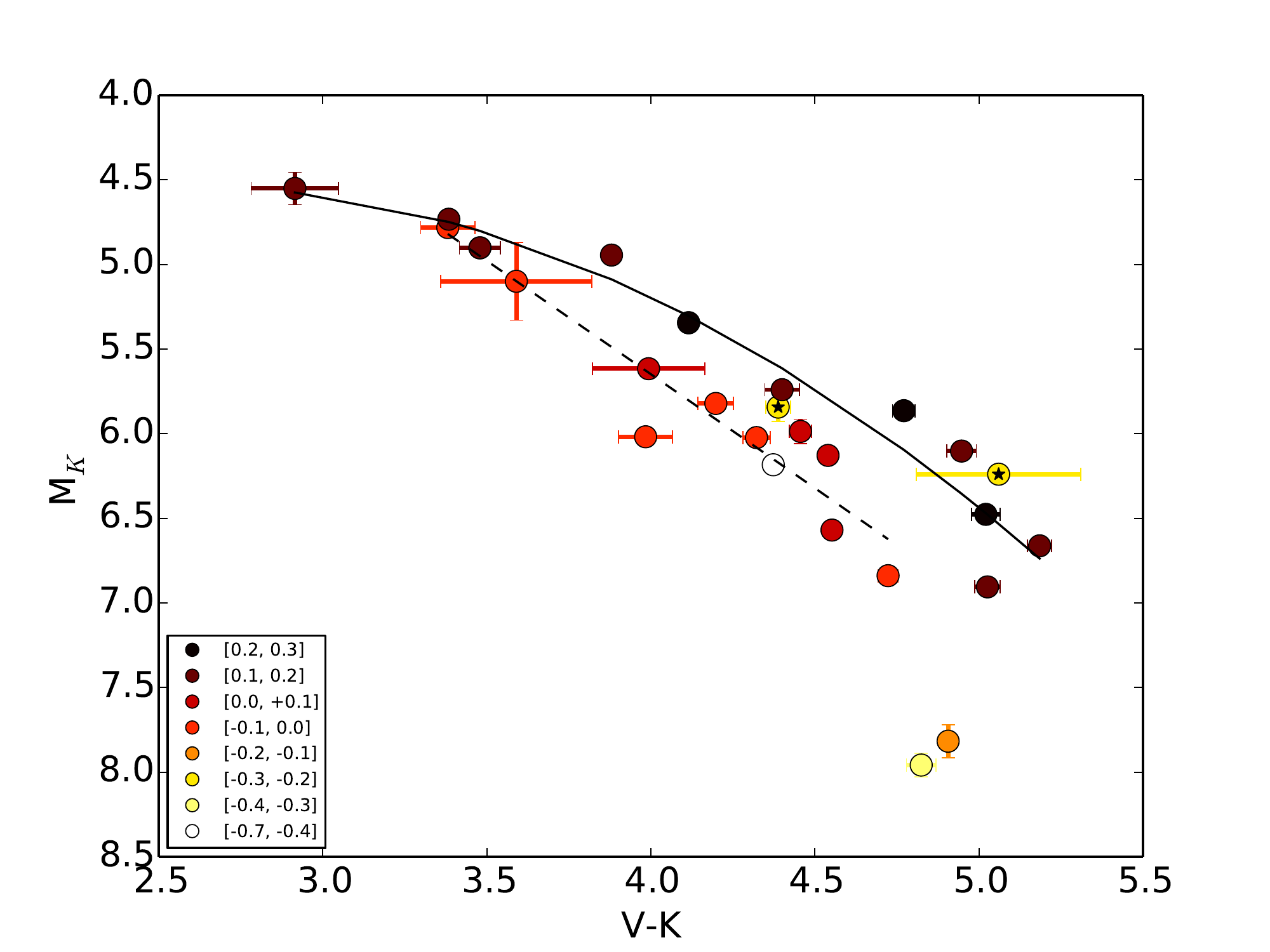}
\caption{Distribution of each targets position in color-magnitude diagrams using the three different infrared magnitudes from 2MASS. Targets include stars both from this paper and from L16. The symbol colors show our metallicities determined from high-resolution infrared spectroscopy.
\label{CMD} }
\end{figure}

\subsection{Color-magnitude diagrams}
\label{sect:CMD}
As our combined samples have a rather good coverage of metallicity and effective temperature we explored the future possibilities for a photometric calibration based on our sample. All current photometric calibrations have been based on the assumption that for a star of given mass the increased amount of metals in the photospheric layer will decrease the overall bolometric luminosity. The increased metallicity also shifts part of the flux into the near-infrared because of the increased opacity in the visible range, mainly due to TiO and VO. In the infrared these effects are predicted to counteract each other, while in the visible the two effects work in the same direction making the $V$ magnitude very sensitive to the stellar metallicity. This results in redder ($V-$infrared band) colors for more metal-rich stars, and therefore a star's position in for example a ($V - K\rm_s$) $-$ $M_{K\rm_s}$ diagram can be used to estimate its metallicity.

In Fig. \ref{CMD} we show the $V-J$, $V-H$ and $V-K\rm_s$ color versus the absolute $K\rm_s$ magnitude, calculated from data in Table~\ref{photometry} assuming negligible interstellar extinction. The color of the symbols shows our determined metallicity using high-resolution infrared spectroscopy. The dashed and solid lines show isometallicity contours derived from second degree polynomial fits to the data points with $-$0.1 $<$ [M/H] $<$ 0.1 and 0.1 $<$ [M/H] $<$ 0.4, respectively. The sub-sample around solar metallicity contains eleven stars and the other sub-sample twelve stars. The more metal-poor stars are too few to attempt any functional fit. 

For targets with bluer colors than approximately $V-J$ $<$ 2.6, $V-H$ $<$ 3.2 and $V-K$ $<$ 3.4 the two functions show little or no separation, hence a photometric calibration can not be defined for this range of colors. However, these stars are K dwarfs and early M dwarfs for which a spectroscopic approach might be applicable. For the redder and cooler targets a photometric calibration seems feasible, with increasing sensitivity to metallicity towards redder colors.

Two stars, GJ~228 and GJ~872B, stand out in Fig.~\ref{CMD} regarding their metallicities. The are marked with a star inside the data point, and we have reason to suspect that their photometry may be incorrect. GJ~228 was found in the Washington Double Star Catalog \citep{Mason2001} with a companion at 0$\farcs$6 separation with a $V$ magnitude of 12.65. Because of the small separation the flux from the companion might have affected the photometry. It may also have influenced the strength of the lines in our observed spectra, even if we do not see any doubling of any spectral lines. For GJ~872B, the plots themselves may be the strongest indication that the photometry is incorrect. GJ~872B is located to the right of, or among, the most metal-rich targets, while we determined a metallicity of $-$0.21~dex. To evaluate whether our metallicity or the photometry is incorrect we compared our metallicity with previous spectroscopic work of its warmer companion, GJ~872A, with spectral type F6V. Since all published values are sub-solar and in good agreement with our value, for example $-$0.22~$\pm$~0.03 \citep{Valenti2005} and  $-$0.24~$\pm$~0.07 \citep{Heiter2003}, we conclude that the photometry is most likely incorrect.

\subsection{Comparison with previous spectroscopic and photometric calibrations}
\label{sect:previous}
\begin{figure*}
	\subfloat[Bonfils et al. (2005)]{\includegraphics[width=0.33\textwidth]{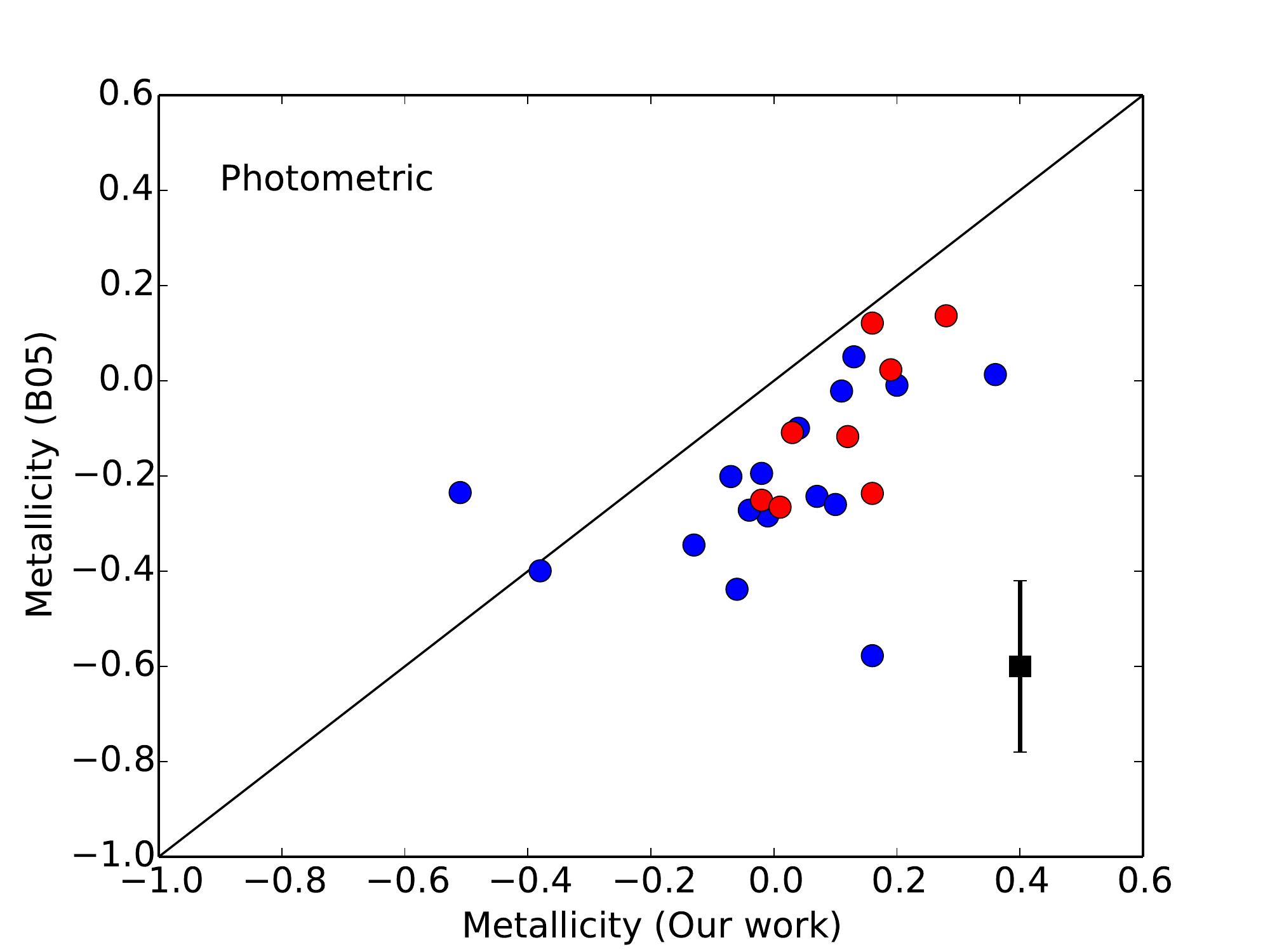}}
	\subfloat[Johnson \& Apps (2009)]{\includegraphics[width=0.33\textwidth]{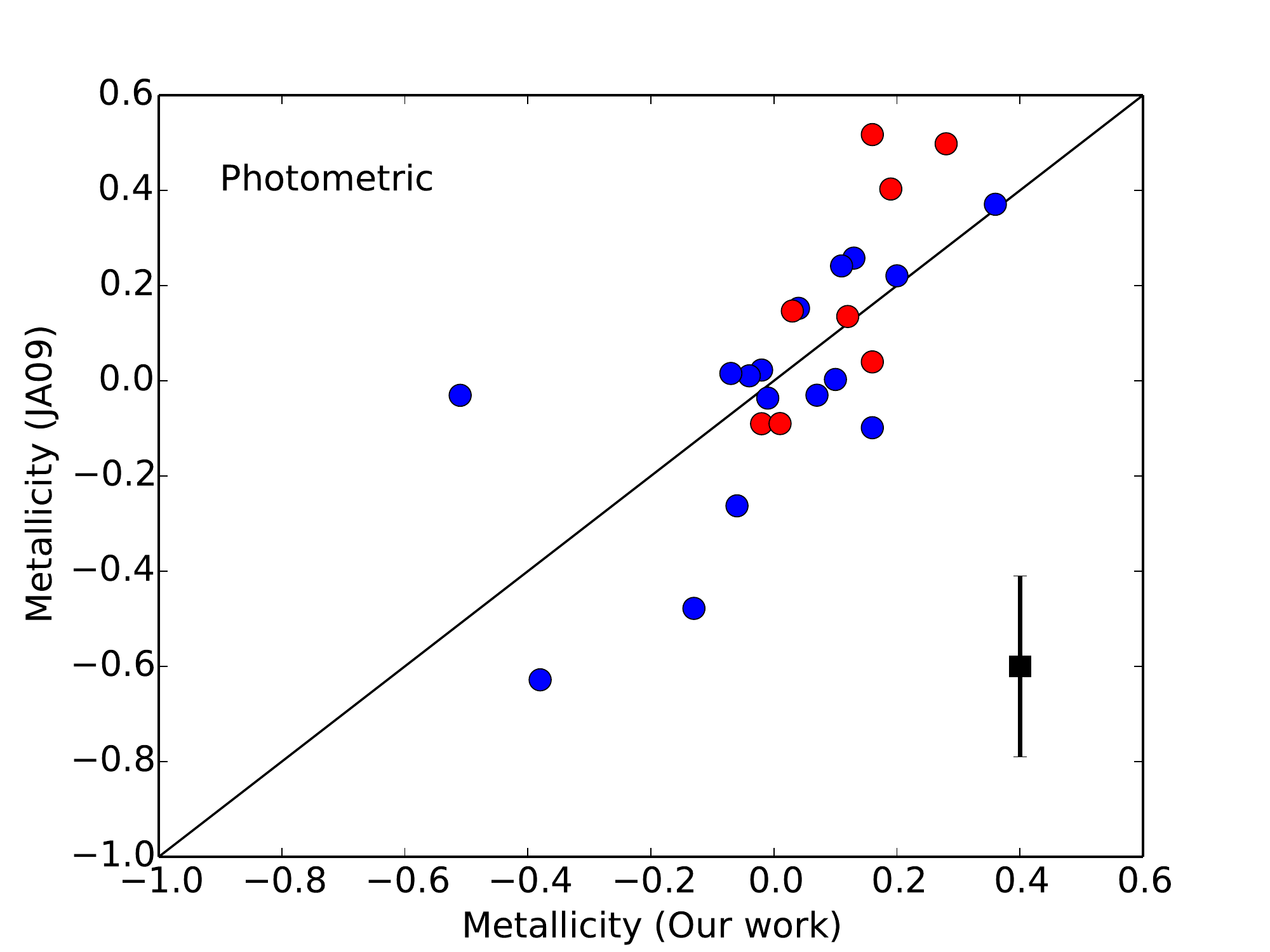}}
	\subfloat[Schlaufman \& Laughlin (2010)]{\includegraphics[width=0.33\textwidth]{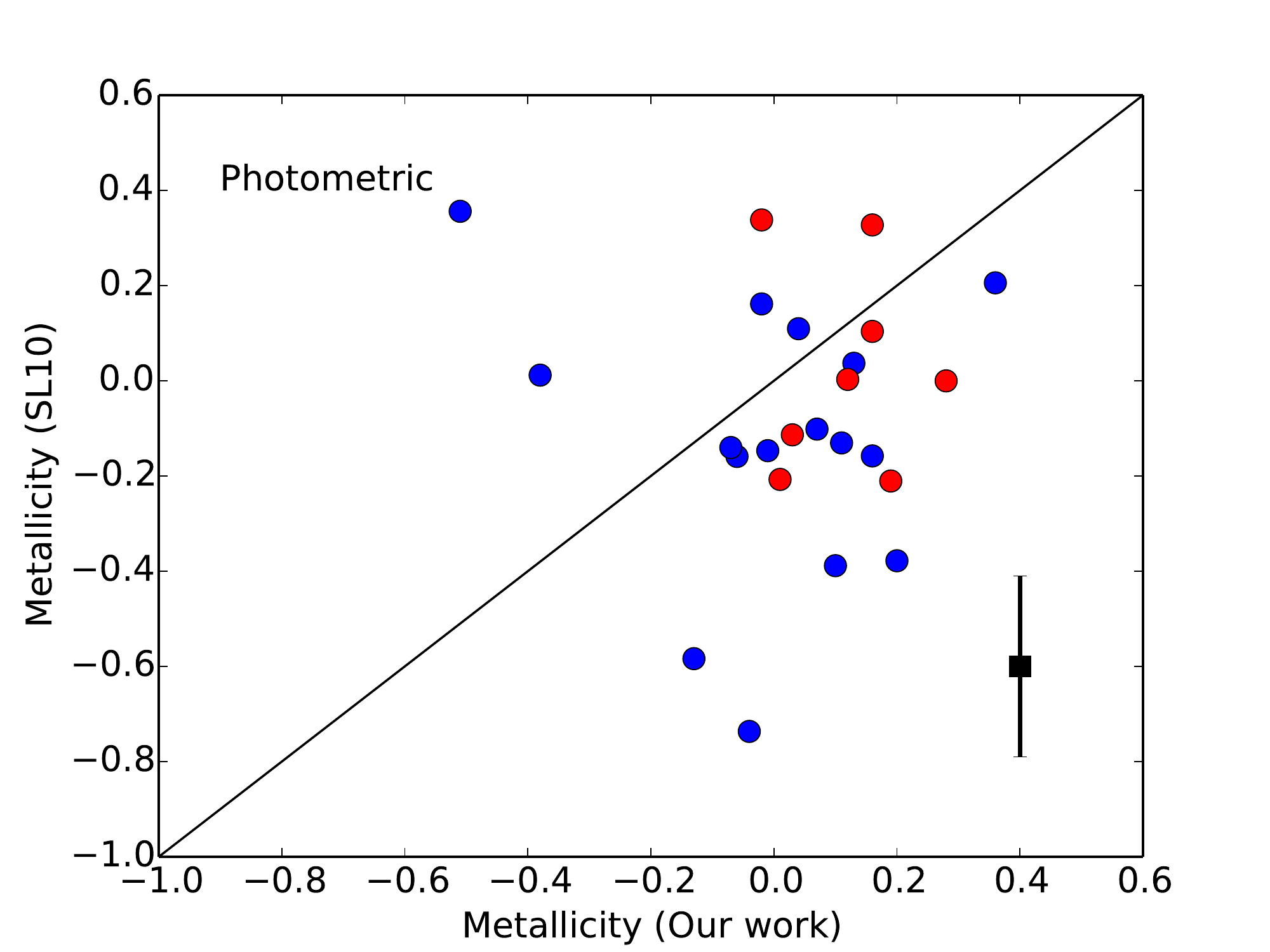}}\\
	\subfloat[Neves et al. (2012)]{\includegraphics[width=0.33\textwidth]{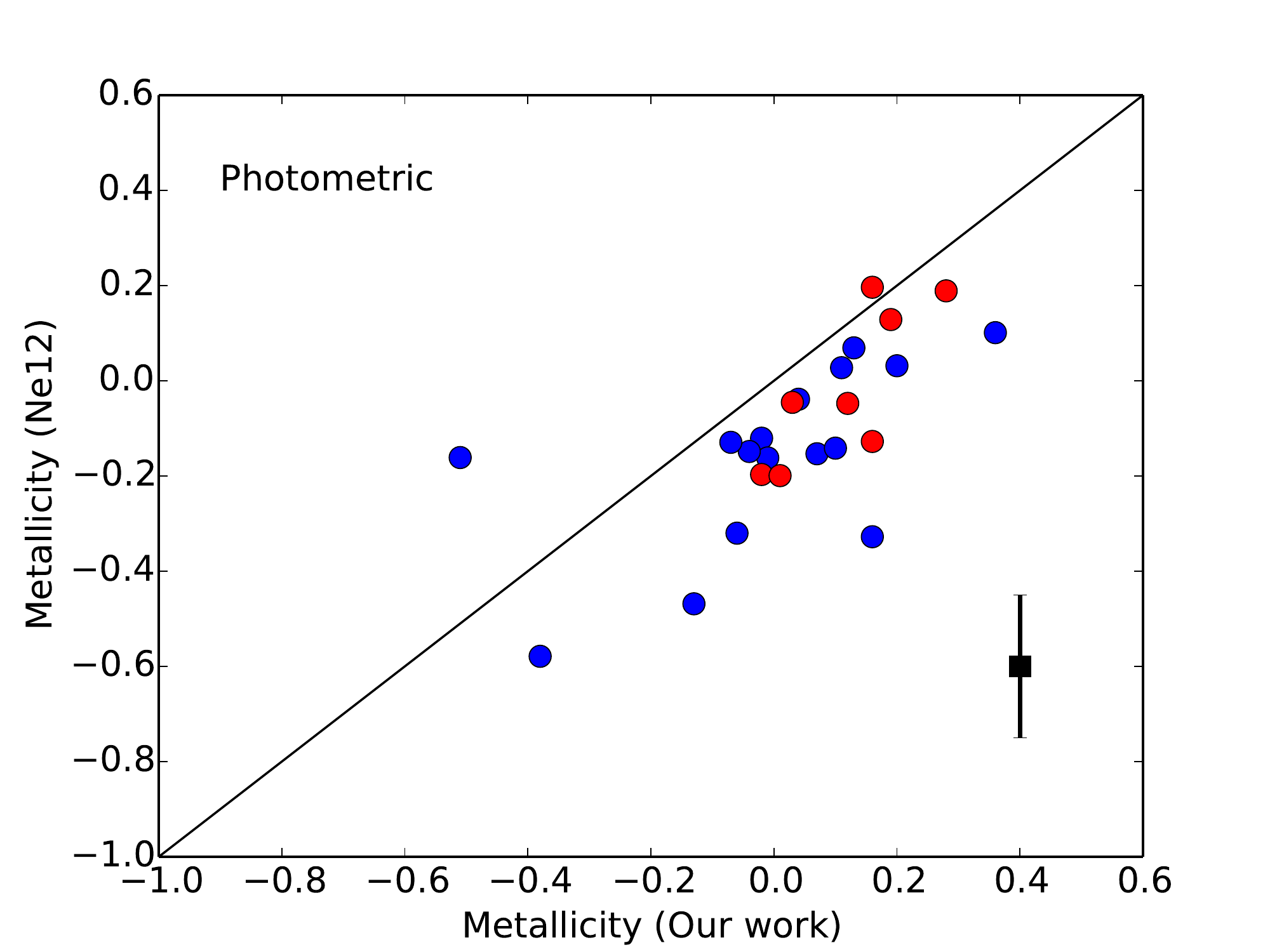}}
	\subfloat[Johnson et al. (2012)]{\includegraphics[width=0.33\textwidth]{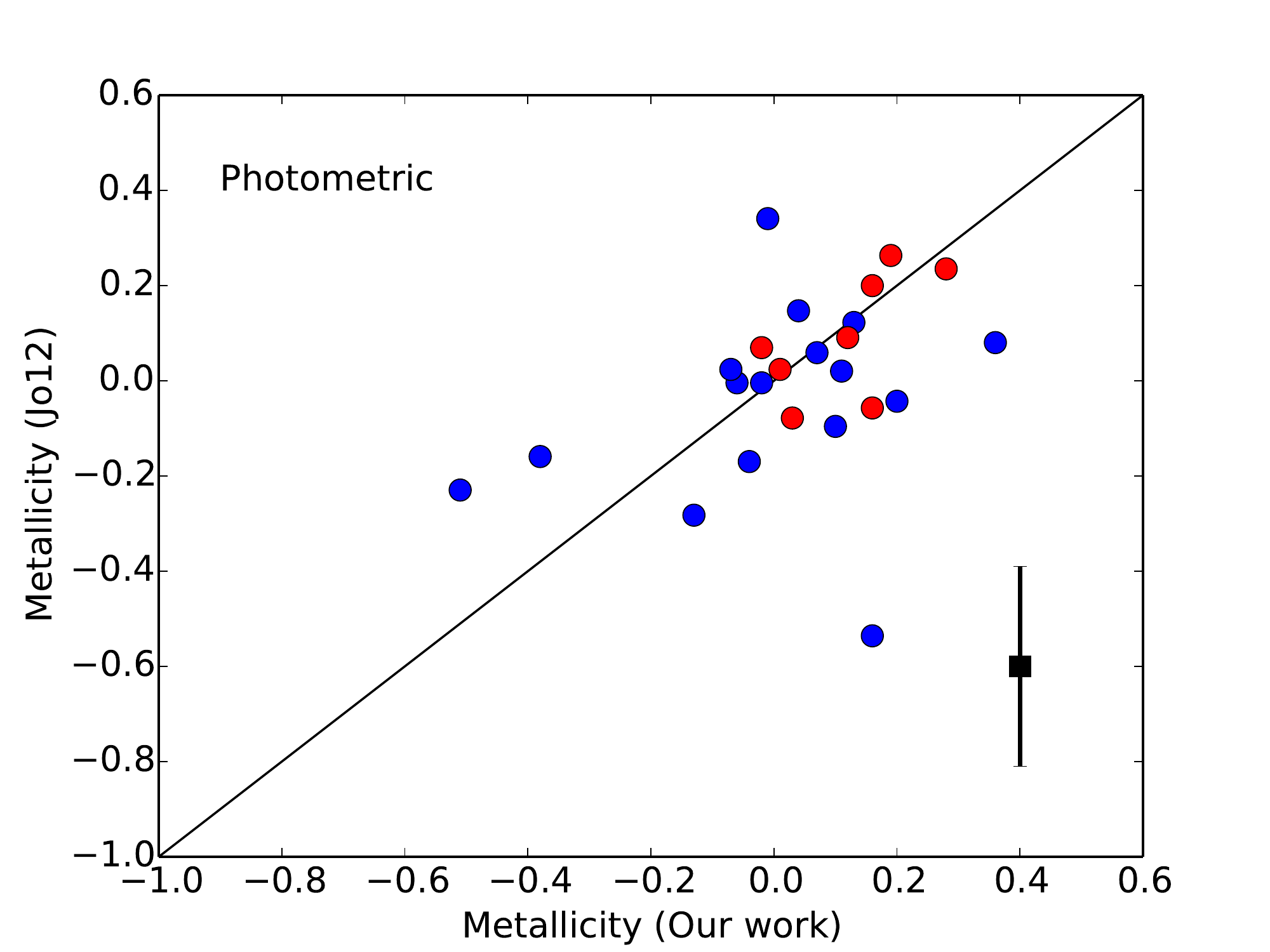}}
	\subfloat[Rojas-Ayala et al. (2012)]{\includegraphics[width=0.33\textwidth]{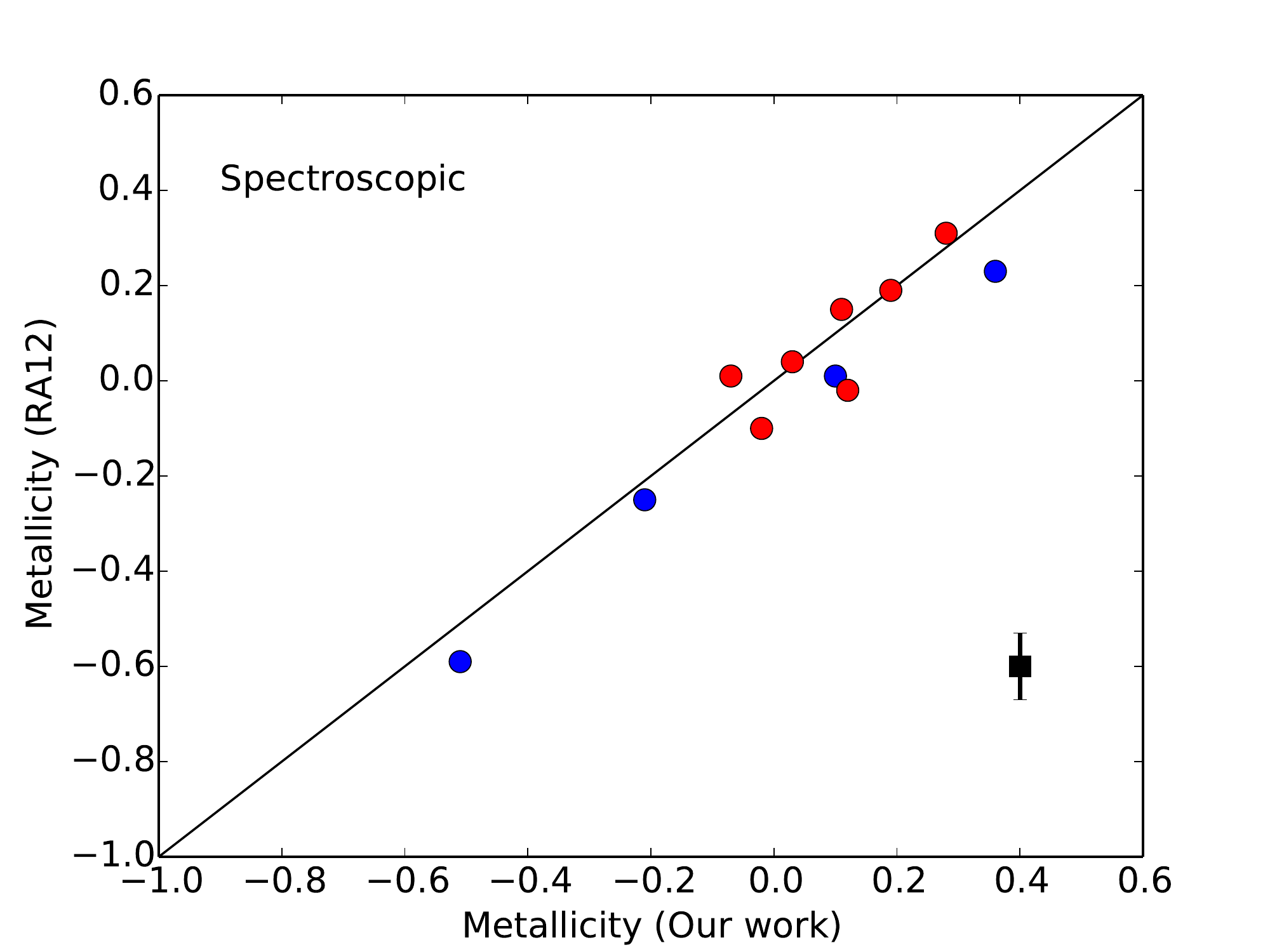}}\\
	\subfloat[Santos et al. (2013)]{\includegraphics[width=0.33\textwidth]{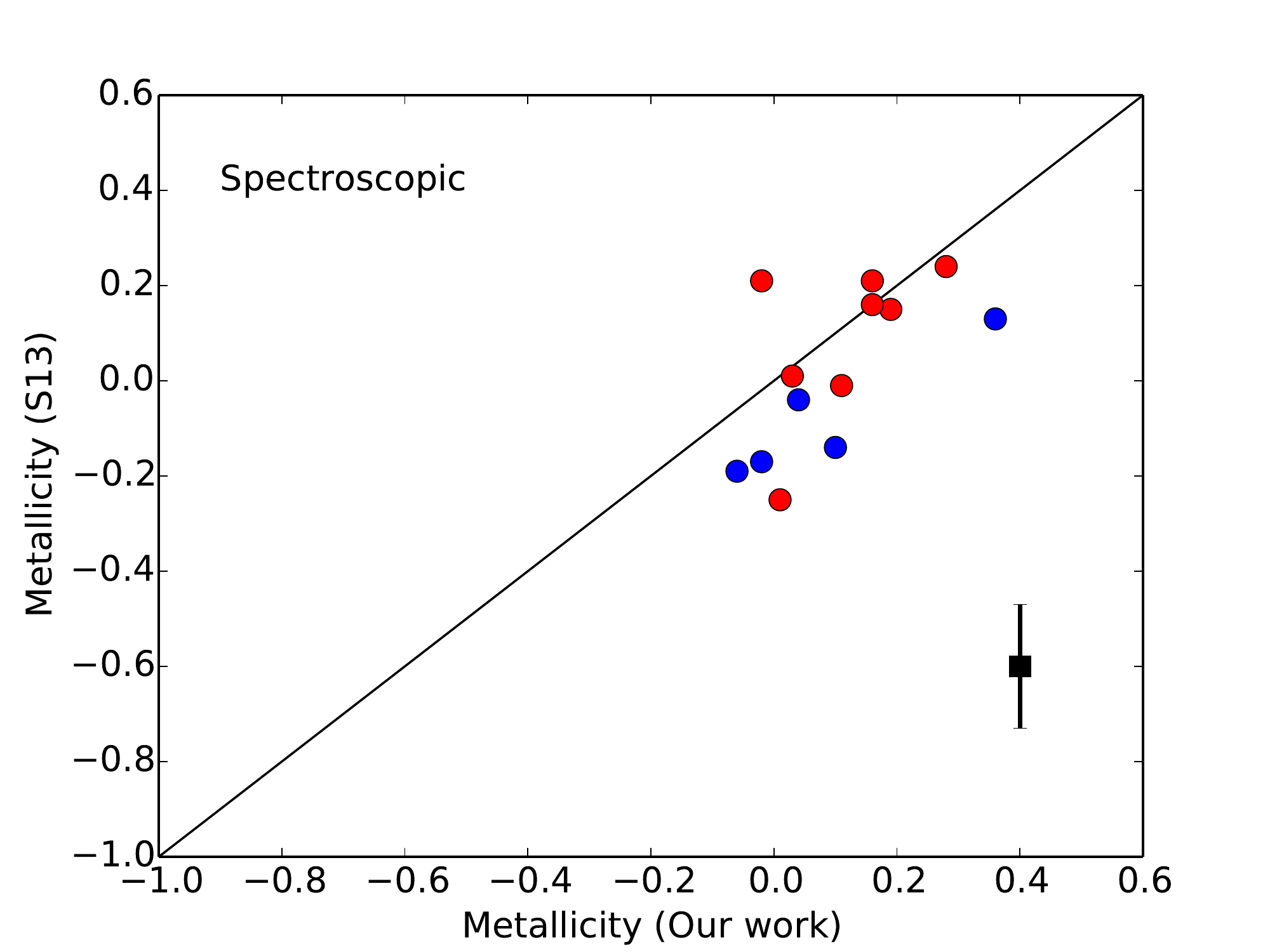}}
	\subfloat[Mann et al. (2015)]{\includegraphics[width=0.33\textwidth]{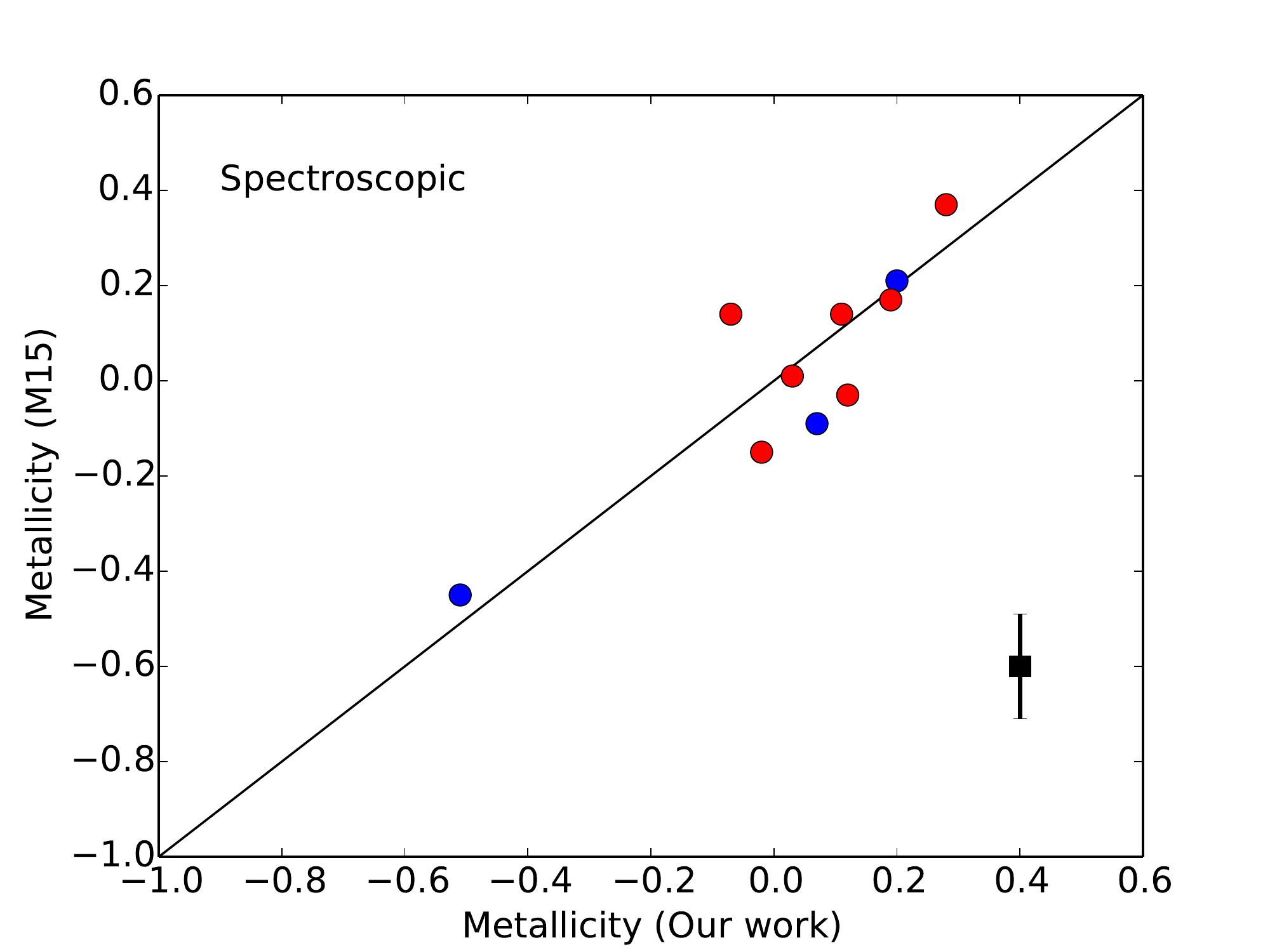}}
	\caption{Comparison of the metallicity values determined by five photomentric calibrations \citep{Bonfils2005, Schlaufman2010, Neves2012, Johnson2009, Johnson2012}, and relevant spectroscopic work (\citealt{Rojas-Ayala2012, Santos2013}, M15). The black lines show a one-to-one relationship, the blue points the metallicity values from this paper and the red points published results from L16. The numerical value of the average difference of our result compared to each calibration is given in the text in Sect.~\ref{sect:previous}. The error bars show the standard deviation of the differences between the metallicities of each of the literature methods and our work. }
\label{comparison} 
\end{figure*}

For a few targets there are metallicity values published using other spectroscopic approaches at lower spectral resolution or at optical wavelengths. We note that all metallicity values from other studies in this section are the iron abundances [Fe/H], while our results are the overall metallicity [M/H]. RA12 used the equivalent widths of the Na~I doublet and Ca~I triplet in the $K$-band. In this wavelength region water forms a pseudo continuum, and the effects of different temperatures were accounted for using their H$_2$O-K2 index. M15 used the metallicity sensitive features found in \citet{Mann2013}, together with different temperature indices: H$_2$O-K2 (RA12), H$_2$O-H \citep{Terrien2012} and H$_2$O-J index \citep{Mann2013}. Lastly we compared our values with those by \citet{Santos2013}, who determined the metallicity from equivalent widths of Fe~I and Fe~II lines in optical spectra. All spectroscopic works are in relatively good agreement with our values, as can be seen in Fig.~\ref{comparison}f-h. The mean differences between the other works and our results are $-$0.04 ($\sigma$~=~0.07) for RA12, $-$0.08 ($\sigma$~=~0.13) for S13, and $-$0.01 ($\sigma$~=~0.11) for M15.

As several of the targets have no previous spectroscopic metallicity estimate we compared our results with photometric calibrations (B05:~\citealt{Bonfils2005}, JA09:~\citealt{Johnson2009}, SL10:~\citealt{Schlaufman2010}, N12:~\citealt{Neves2012}, J12:~\citealt{Johnson2012}), see Fig. \ref{comparison}a-e. However, because of the earlier discussed probable inconsistencies with the photometry for GJ~228 and GJ~872B we excluded them from the comparison with the photometric studies. The calibration by \citet{Bonfils2005} mainly gives lower values than our determined values. The average difference between this calibration and our high-resolution analysis is $-$0.21~dex. Also the values by \citet{Schlaufman2010} and \citet{Neves2012} are on average lower ($-$0.12 and $-$0.15, respectively). The calibration by \citet{Johnson2009} as well as their revised calibration \citep{Johnson2012} agree better with our metallicities (+0.02 and $-$0.04, respectively). However, the spread of the differences within all empirical calibrations is significant with standard deviations of $\sigma_{\rm B05}$~=~0.18, $\sigma_{\rm JA09}$~=~0.19,  $\sigma_{\rm SL10}$~=~0.19, $\sigma_{\rm N12}$~=~0.15 and $\sigma_{\rm J12}$~=~0.21.

\section{Summary}
In this paper we have determined the stellar parameters of sixteen very low-mass stars using high-resolution infrared spectroscopy with the goal to determine accurate metallicities for M dwarfs within an increased parameter range. We especially aimed at more metal-poor stars compared to L16, in order to achieve a combined sample that can be used for a future photometric calibration. Based on the color-magnitude diagrams our sample seems to be sufficient for a calibration for M dwarfs with metallicities between $-$0.1 to 0.3 dex, while even the combined sample does not have enough M dwarfs to make a reliable calibration for more metal-poor stars. 

We further confirm that all available empirical photometric metallicity calibrations give an unsatisfactorily large spread compared to our result using high-resolution spectroscopy. Compared to low-resolution spectroscopic work we found a better agreement, with the best agreement with RA12 and M15. Even if this paper contains mainly warmer targets than L16, for the few targets below 3575~K we confirmed that the use of FeH lines can be used to determine the effective temperature. 

Our combined sample contains twelve planet hosts. Even if the majority of them are metal-enhanced compared to the Sun we can not say anything regarding the nature of a planet-host metallicity correlation for M dwarfs because our sample is significantly affected by selection effects. The sample in L16 was mainly selected to include metal-rich stars and planet hosts, while the sample in this paper was selected to include mainly metal-poor stars. 

We point out that three of the stars analyzed in L16 and in this paper have been selected as M-dwarf benchmark stars for the Gaia-ESO public spectroscopic survey \citep[Sect. 4.2 in][]{Pancino2017}. These are GJ~436, GJ~581 (L16), and GJ~880 (this paper). Numerous high-resolution spectra have been obtained for these stars with the Gaia-ESO setups, and they are used to test and validate the analysis methods within Gaia-ESO. For all of them angular diameter measurements and bolometric flux measurements are available, which give the effective temperature independently of photometry or spectroscopy \citep[Table 5 in][]{Pancino2017}. The $T_{\rm eff}$ values are $\sim$3400~K for GJ~436 and GJ~581, similar to the values derived in L16, and 3710$\pm$40~K for GJ~880, which is equal to the spectroscopic value of 3720$\pm$60 K determined by M15 that we used in this work. We recommend that any future observational project on M dwarfs should include some of the benchmark stars listed in \citet{Pancino2017} among the targets, if possible, for easier cross-validation of different projects. \\

\begin{acknowledgements}
U.H. acknowledges support from the Swedish National Space Board (SNSB/Rymdstyrelsen). This work has made use of the VALD database, operated at Uppsala University, the Institute of Astronomy RAS in Moscow, and the University of Vienna. This work has made use of the NSO/Kitt Peak FTS data, produced by NSF/NOAO. This work has made use of data from the European Space Agency (ESA) mission \emph{Gaia} (\url{http://www.cosmos.esa.int/gaia}), processed by the  \emph{Gaia} Data Processing and Analysis Consortium (DPAC, \url{http://www.cosmos.esa.int/web/gaia/dpac/consortium}). Funding for the DPAC has been provided by national institutions, in particular the institutions participating in the \emph{Gaia} Multilateral Agreement. We also want to thank our anonymous referee for useful comments that helped improve the paper.

\end{acknowledgements}

\bibliographystyle{aa}
\bibliography{biblio}

\newpage

\onecolumn
\Online

\begin{appendix}
\section{Additional FeH plots} \label{app:linelist}

\begin{figure}[h!]
\label{something}
%\begin{center}
\includegraphics[width=0.33\textwidth]{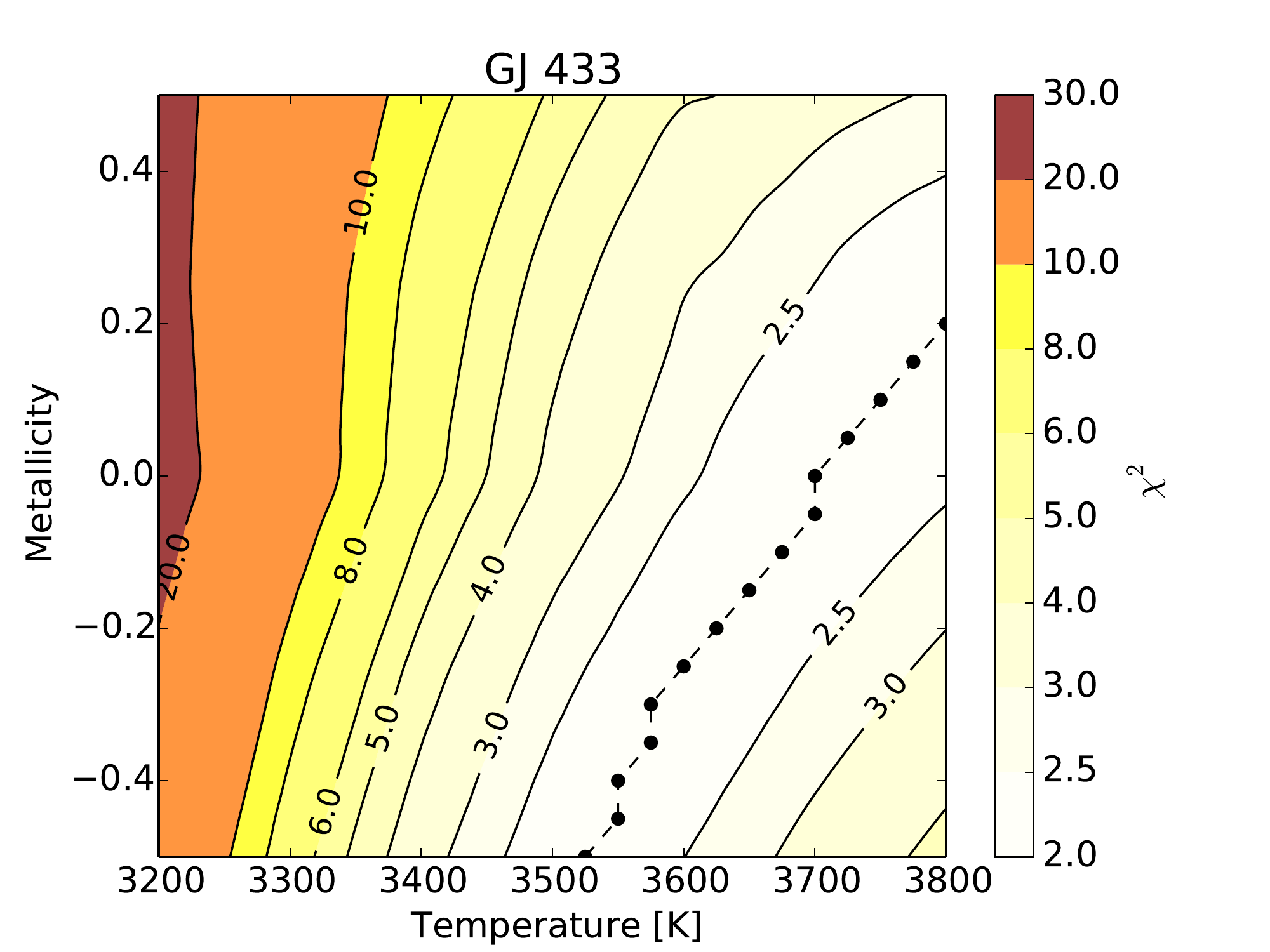}
\includegraphics[width=0.33\textwidth]{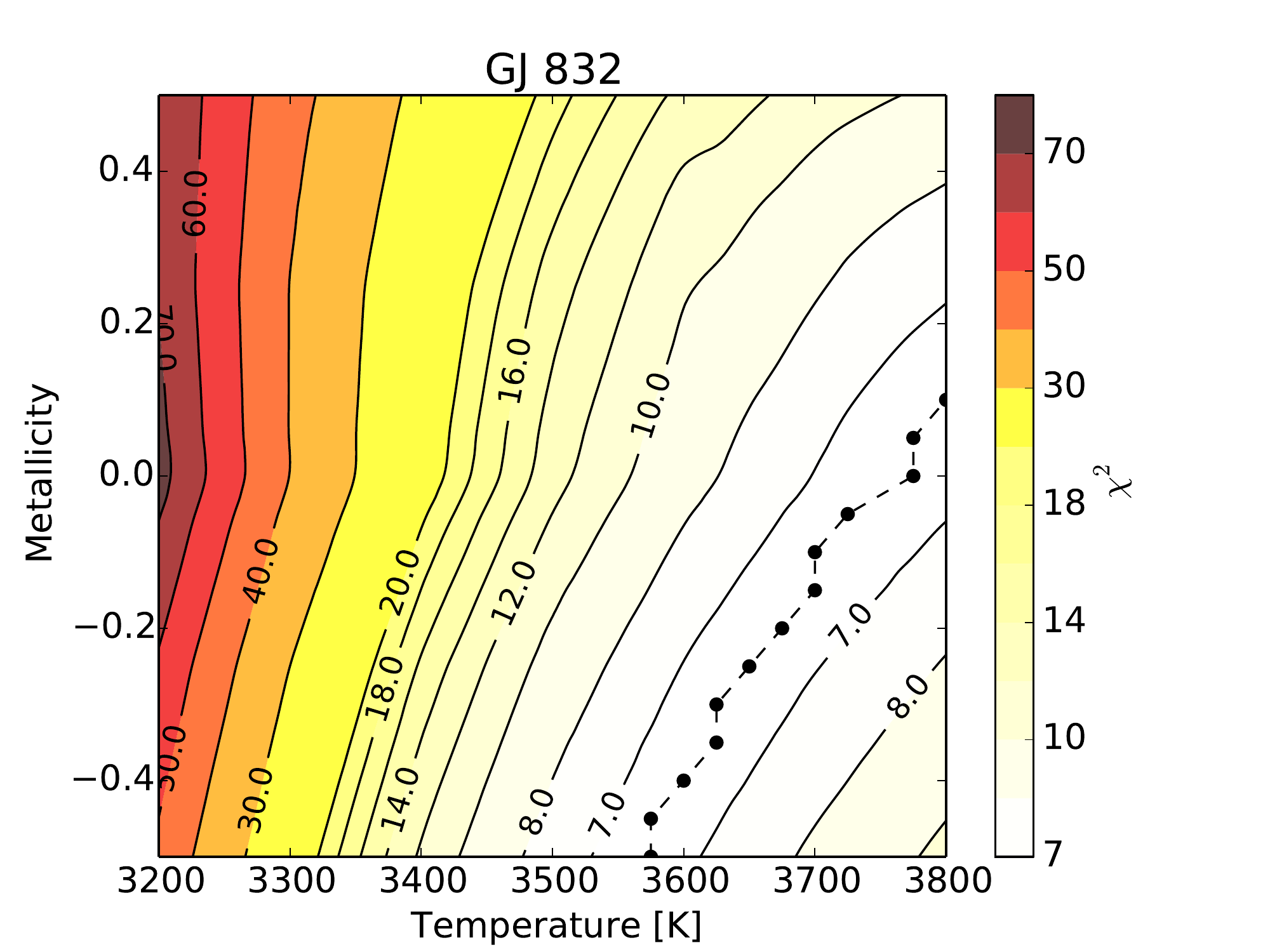}
\includegraphics[width=0.33\textwidth]{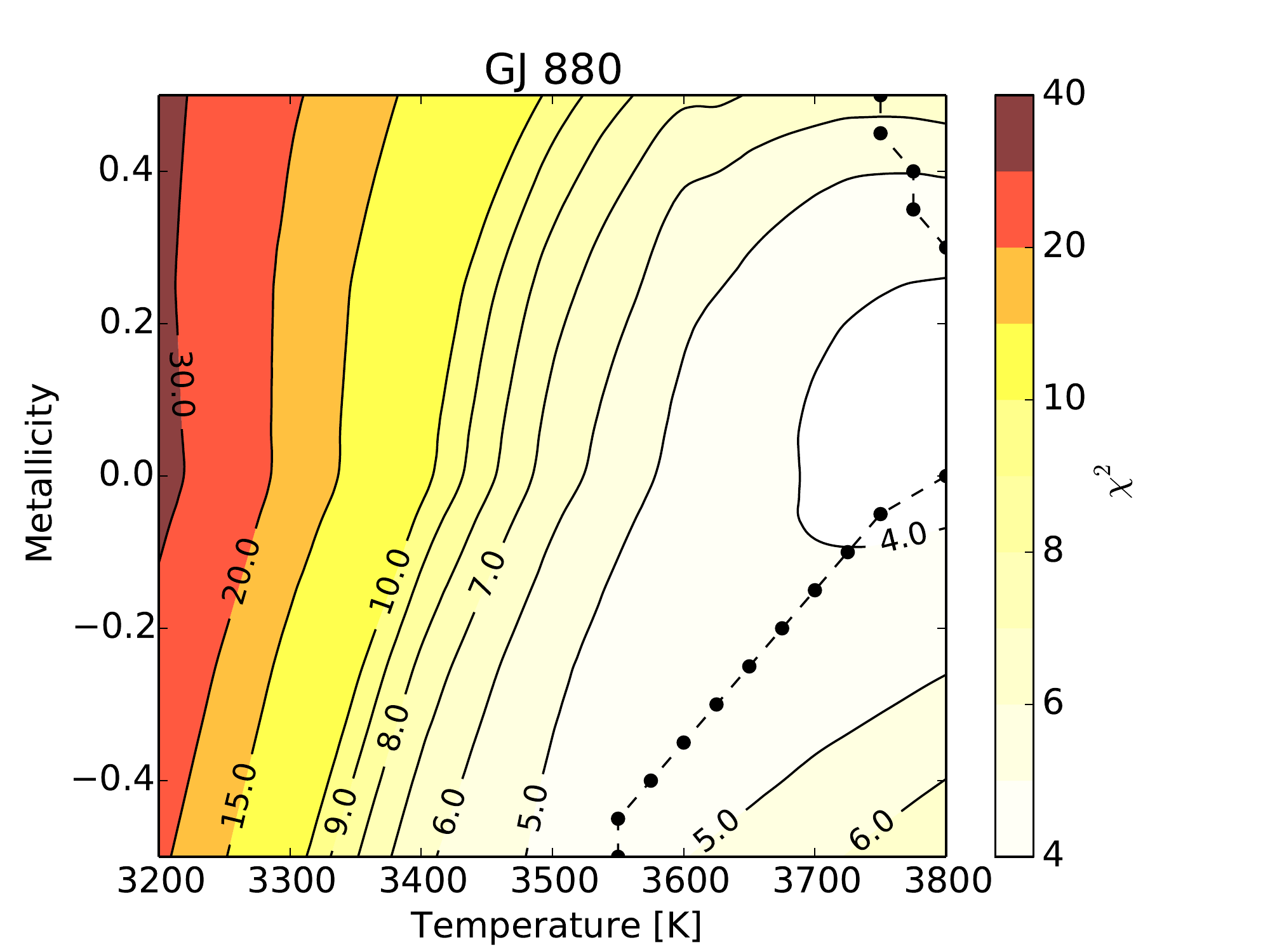}
\includegraphics[width=0.33\textwidth]{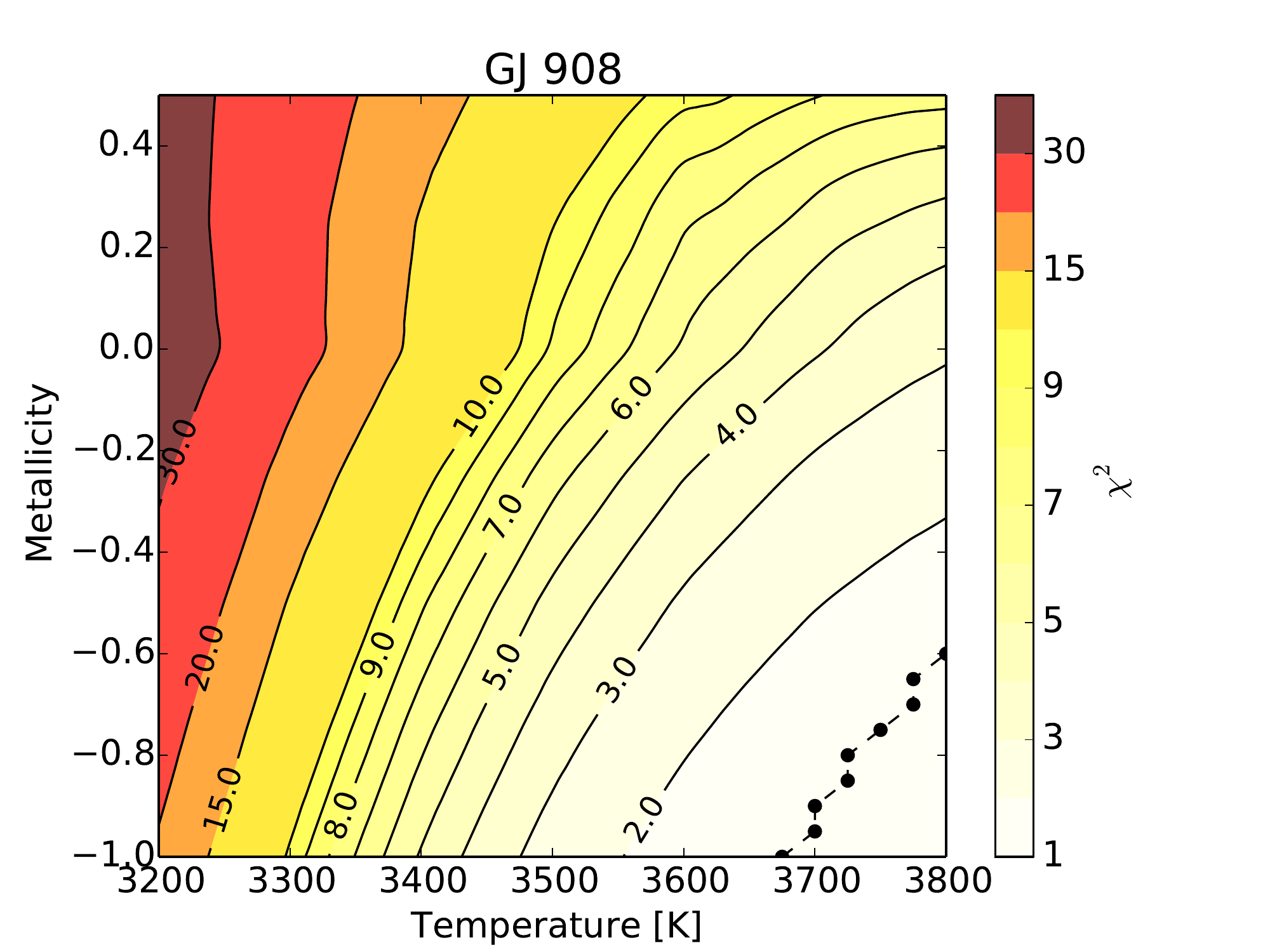}
%\end{center}
\caption{Demonstration of FeH line strength dependency on the effective temperature and overall metallicity. The contour plots in this figure show the calculated $\chi^2$ of the fit based on a grid of metallicity and effective temperature of the stars with spectral type M1.5. The connected dots indicate the temperature with the minimum $\chi^2$ for each step in metallicity. Note that the range for metallically is larger for GJ~908 than the remaining targets.}
\end{figure}

\end{appendix}

\end{document}